\documentclass[pre,preprint,showpacs,superscriptaddress]{revtex4}
\usepackage{graphicx}
\usepackage{amsmath}
\usepackage{setspace} 

\usepackage[colorlinks=true,citecolor=black,linkcolor=black,bookmarksnumbered=true]{hyperref}
\hypersetup{%
  pdftitle = {Stochastic simulations of cargo transport},
  pdfkeywords = {molecular motors, stochastic processes, Stokesian dynamics},
  pdfauthor = {Christian Korn, Stefan Klumpp, Reinhard Lipowsky, Ulrich Schwarz}
}

\def\eq#1{Eq.~(\ref{#1})}
\def\fig#1{Fig.~\ref{#1}}
\def\tab#1{Tab.~\ref{#1}}
\def\sec#1{Sec.~\ref{#1}}

\newcommand{\boldv}[1]{{\mathbf #1}}

\newcommand{\ha}{\hspace{1cm}}

\newcommand{\grad}{{\mathbf \nabla}}

\newcommand{\mean}[1]{\langle #1 \rangle}

\newcommand{\kk}{\kappa} 
\newcommand{\Nb}{N_m}
\newcommand{\Nbmax}{N_{m,max}}
\newcommand{\mNb}{{N}_b}
\newcommand{\N}{N}
\newcommand{\dxb}{\mean{\Delta x_b}}

\begin{document}

\title{Stochastic simulations of cargo transport\\
by processive molecular motors}

\author{Christian B. Korn}
\affiliation{University of Heidelberg, Bioquant 0013,
Im Neuenheimer Feld 267, D-69120 Heidelberg, Germany}
\author{Stefan Klumpp}
\affiliation{Center for Theoretical Biological Physics, University of
  California at San Diego, 9500 Gilman Drive, La Jolla, CA 92093-0374, USA}
\affiliation{Max Planck Institute for Colloids and Interfaces, Science Park Golm,
  D-14424 Potsdam, Germany}
\author{Reinhard Lipowsky}
\affiliation{Max Planck Institute for Colloids and Interfaces, Science Park Golm,
  D-14424 Potsdam, Germany}
\author{Ulrich S. Schwarz}
\affiliation{University of Heidelberg, Bioquant 0013,
Im Neuenheimer Feld 267, D-69120 Heidelberg, Germany}
\affiliation{University of Heidelberg, Institute for 
Theoretical Physics, Philosophenweg 19, D-69120 Heidelberg, Germany}

\begin{abstract}
  We use stochastic computer simulations to study the transport of a
  spherical cargo particle along a microtubule-like track on a planar
  substrate by several kinesin-like processive motors. Our newly
  developed adhesive motor dynamics algorithm combines the numerical
  integration of a Langevin equation for the motion of a sphere with
  kinetic rules for the molecular motors. The Langevin
  part includes diffusive motion, the action of the pulling motors,
  and hydrodynamic interactions between sphere and wall. The kinetic
  rules for the motors include binding to and unbinding from the
  filament as well as active motor steps. We find that the simulated
  mean transport length increases exponentially with the number of
  bound motors, in good agreement with earlier results. The number of
  motors in binding range to the motor track fluctuates in time with a
  Poissonian distribution, both for springs and cables being used as
  models for the linker mechanics. Cooperativity in the sense of equal
  load sharing only occurs for high values for viscosity and
  attachment time.
\end{abstract}

\pacs{82.39.-k,87.10.Mn,87.15.hj}

\maketitle

\section{Introduction}

Molecular motors play a key role for the generation of movement and
force in cellular systems \cite{b:howa01}. In general there are two
fundamentally different classes of molecular motors. Non-processive
motors like myosin II motors in skeletal muscle bind to their tracks
only for relatively short times. In order to generate movement and
force, they therefore have to operate in sufficiently large numbers.
Processive motors like kinesin remain attached to their tracks for a
relatively long time and therefore are able to transport cargo over
reasonable distances. Indeed processive cytoskeletal motors
predominantly act as transport engines for cargo particles, including
vesicles, small organelles, nuclei or viruses.  For example, kinesin-1
motors make an average of 100 steps of size 8 nm along a microtubule
before detaching from the microtubule \cite{coy:99,block:99}, and
therefore reach typical run lengths of micrometers.

However, for intracellular transport even processive motors tend to
function in ensembles of several motors, with typical motor numbers in
the range of 1-10 \cite{Gross__Shubeita2007}. The cooperation of
several motors is required, for example, when processes like extrusion
of lipid tethers require a certain level of force that exceeds the
force generated by a single motor \cite{c:kost03,c:ledu04}.  The
cooperative action of several processive motors is also required to
achieve sufficiently long run length for cargo transport
\cite{klumpp:05}, as transport distances within cells are typically of
the order of the cell size, larger than the micron single motor run
length. In this context the most prominent example is axonal
transport, as axons can extend over many centimeters
\cite{Goldstein_Yang2000,hill:04}. Another level of complexity of
transport within cells is obtained by the simultaneous presence of
different motor species on the same cargo, which can lead to
bidirectional movements and switching between different types of
tracks \cite{Gross2004,Mueller__Lipowsky2008}, and by exchange of
components of the motor complex with the cytoplasm.

Cargo transport by molecular motors can be reconstituted \emph{in
  vitro} using so-called bead assays in which motor molecules are
firmly attached to spherical beads that flow in aqueous solution in a
chamber. On the bottom wall of the chamber microtubules are mounted
along which the beads can be transported
\cite{b:howa01,Block__Schnapp1990,boehm:01,beeg:08}. This assay has
been used extensively to study transport by a single motor over the
last decade \cite{Block__Schnapp1990}, but recently several groups
have adapted it for the quantitative characterization of transport by
several motors \cite{Vershinin__Gross2007,beeg:08}. If several motors
on the cargo can bind to the microtubule, then the transport process
continues until all motors simultaneously unbind from the microtubule.
Based on a theoretical model for cooperative transport by several
processive motors, it was recently predicted that the mean transport
distance increases essentially exponentially with the number of
available motors \cite{klumpp:05}. Indeed these predictions are in
good agreement with experimental data \cite{beeg:08,rogers09}.
However, both the theoretical approach and the experiments do not
allow us to investigate the details of this transport process.
A major limitation of the bead assays for transport
by several motors is that the number of motors per bead varies
from bead to bead and that only the average number of motors per bead is
known \cite{Vershinin__Gross2007,beeg:08}. In addition, even if
the number of motors on the bead was known, the number of motors
in binding range would still be a fluctuating quantity. Recently two kinesin
motors have been elastically coupled by a DNA scaffold and the
resulting transport has been analyzed in quantitative detail
\cite{rogers09}. However, it is experimentally very challenging to
extend this approach to higher numbers of motors.

One key property of transport by molecular motors is the load force
dependence of the transport velocity.  For transport by single
kinesins, the velocity decreases approximately linearly with
increasing load and stalls at a load of about 6~pN \cite{block:99}.
Thus, when the cargo has to be transported against a large force, the
speed of a single motor is slowed down. However, if several motors
simultaneously pull the cargo, they could share the total load. This
cooperativity lets them pull the cargo faster. Assuming equal load
sharing, one can show that in the limit of large viscous load
force the cargo velocity is expected to be proportional to the number
of pulling motors \cite{klumpp:05}. Indeed, this is one explanation
that was proposed by Hill et al. to give plausibility to their results
from \emph{in vivo} experiments, which showed that motor-pulled
vesicles move at speeds of integer multiples of a certain velocity
\cite{hill:04}. In general, however, one expects that the total load
is not equally shared by the set of pulling motors. The force
experienced by each motor will depend on its relative position along
the track and can be expected to fluctuate due to the stochasticity of
the motor steps \cite{Kunwar__Gross2008}. In addition, for a spherical
cargo particle curvature effects are expected to play a role of the
way force is transmitted to the different motors. Because of its small
size, the cargo particle is perpetually subject to thermal
fluctuations. This diffusive particle motion is also expected to
affect the load distribution and depends on the exact height of the
sphere above the wall due to the hydrodynamic interactions.

In order to investigate these effects, here we introduce an algorithm
that allows us to simulate the transport of a spherical particle by
kinesin-like motors along a straight filament that is mounted to a
plain wall. Binding and unbinding of the motor to the filament can be
described in the same theoretical framework as the reaction dynamics
of receptor-ligand bonds \cite{lauffenburger,uss:erdm04a,erdmann:04b}.
Similarly as receptors bind very specifically to certain ligands,
conventional kinesin binds only to certain sites on the microtubule.
Thus from the theoretical point of view a spherical particle covered
with motors binding to tracks on the substrate is equivalent to a
receptor-covered cell binding to a ligand-covered substrate. This
situation is reminescent of rolling adhesion, the phenomenon that in
the vasculature different cell types (mainly white blood cells, but
also cancer cells, stem cells or malaria-infected red blood cells)
bind to the vessel walls under transport conditions \cite{c:spri94}.
Different approaches have been developed to understand the combination
of transport and receptor-ligand kinetics in rolling adhesion. Among
these Hammer and coworkers developed an algorithm that combines
hydrodynamic interactions with reaction kinetics for receptor-ligand
bonds \cite{hammer:92}. Recently, we introduced a new version of this
algorithm that also includes diffusive motion of the spherical
particle \cite{korn:08a}. Here, we further extend our algorithm to
include the active stepping of motors (\emph{adhesive motor
  dynamics}). Simulation experiments with this algorithm provide
access to experimentally hidden observables like the number of
actually pulling motors, the relative position of the motors to each
other, and load distributions.  The influence of thermal fluctuations
on the motion of the cargo particle is also influenced by the
properties of the molecules that link the cargo to the microtubule. In
our simulations various polymer models can be implemented and their
influence can be tested directly. In general our method makes it
possible to probe the effects of various microscopic models for motor
mechanics on macroscopic observables that are directly accessible to
experiments.

The organization of the article is as follows. In the first part,
\sec{sec:method}, we explain our model in detail. This is based on a
Langevin equation that allows us to calculate the position and
orientation of a spherical cargo particle as a function of time. In
addition, we include rules that model the reaction kinetics of the
molecular motors being attached to the cargo and comment on the
different kinds of friction involved. We then explain how theoretical
results for the dependence of the mean run length of a cargo particle
on the number of available motors previously obtained in the framework
of a master equations can be compared to the situation where only the
total number of motors attached to the cargo sphere is known. We also
briefly comment on the implementation of our simulations. In the
results part, \sec{sec:results}, we first measure the mean run length
and the mean number of pulling motors at low viscous drag and find
good agreement with earlier results. We then present measurements of
quantities that are not accessible in earlier approaches, including
the dynamics of the number of motors on the cargo that are in binding
range to the microtubule. Finally, we consider cargo transport in the
high viscosity regime and investigate how the load is distributed
among the pulling motors. We find that cooperativity by load sharing
strongly depends on appropriate life times of bound motors.  In the
closing part, \sec{sec:discussion}, we discuss to what extend our
simulations connect theoretical modeling with experimental findings.
Furthermore, we give an outlook on further possible applications of
the adhesive motor dynamics algorithm introduced here.

\section{Model and computational methods}
\label{sec:method}

\subsection{Bead dynamics}
\label{sec:stokesian}

In experiments using bead assays for studying the collective transport
behavior of kinesin motors one notes the presence of three very
different length scales, namely the chamber dimension, the bead size
and the molecular dimensions of kinesin and microtubule, respectively.
The chamber dimension is macroscopic.  The typical radius $R$ of beads
in assay chambers is in the micrometer range \cite{boehm:01}. The
kinesin molecules with which they are covered have a resting length
$l_0$ of about 80~nm \cite{schliwa:03}. Kinesins walk along
microtubules which are long hollow cylindrical filaments 
made from 13 parallel protofilaments and with a
diameter of about $h_{MT} = 24$~nm \cite{limberis:01}. Thus the
chamber dimensions are large compared to the bead radius, which in
turn is large compared to the motors and their tracks.  
This separation of length scales allows us to model the 
microtubule as a line of binding sites covering the wall
and means that the dominant hydrodynamic interaction is the
one between the spherical cargo and the wall. For sufficiently
small motor density, this separation of length scales also
implies that we have to consider only one lane of binding sites.
In the following we therefore consider a rigid
sphere of radius $R$ moving above a planar wall with an embedded line
of binding sites as a simple model
system for the cargo transport by molecular motors along a filament.

For small objects like microspheres typical values of the Reynolds
number are much smaller than one and inertia can be neglected
(overdamped regime). Therefore, the hydrodynamic interaction between
the sphere and the wall is described by the Stokes equation.
Throughout this paper we consider vanishing external flow around the
bead. Directional motion of the sphere arises from the pulling forces
exerted by the motors. In addition the bead is subject to thermal
fluctuations that are ubiquitous for microscopic objects. Trajectories
of the bead are therefore described by an appropriate
Langevin equation. For the sake of a concise notation we introduce the
six-dimensional state vector $\mathbf{X}$ which includes both the
three translational and the three rotational degrees of freedom of the
sphere.  The translational degrees of freedom of $\mathbf{X}$ refer to
the center of mass of the sphere with respect to some reference frame
(cf. \fig{fig:setup}). The rotational part of $\mathbf{X}$ denotes the
angles by which the coordinate system fixed to the sphere is rotated
relative to the reference frame \cite{korn:07a}.  Similarly,
$\mathbf{F}$ denotes a combined six-dimensional force/torque vector.

With this notation at hand the appropriate Langevin equation reads \cite{ermak:78,brady:89}
\begin{align}
  \label{langevin-ito}
  \dot{\boldv{X}} = \mathsf{M}\boldv{F}  + k_B T\grad\mathsf{M} + \boldv{g}_t^I.
\end{align}
Here, $\mathsf{M}$ is the position-dependent $6\times 6$ mobility
matrix. As we consider no-slip boundary conditions at the wall,
$\mathsf{M}$ depends on the height of the sphere above the wall in
such a way that the mobility is zero when the sphere touches the wall
\cite{jones:98}. Thus, the hydrodynamic interaction between the sphere
and the wall is completely included in the configuration dependence of
the mobility matrix. The last term in \eq{langevin-ito} is a Gaussian
white noise term with
\begin{align}
  \label{gauss_noise}
  \mean{\boldv{g}_t^I} = 0,\;\;\;\;
  \mean{\boldv{g}_t^I\boldv{g}_{t'}^I} = 2 k_B T \mathsf{M}\delta(t-t'),
\end{align}
with Boltzmann's constant $k_B$.  The second equation represents the
fluctuation-dissipation theorem illustrating that the noise is
\emph{multiplicative} due to the position-dependance of $\mathsf{M}$.
Thus we also have to define in which sense the noise in
\eq{langevin-ito} shall be interpreted.  As usual for physical
processes modeled in the limit of vanishing correlation time we choose
the Stratonovich interpretation \cite{horsthemke:84}. However,
\eq{langevin-ito} is written in the It{\^o} version marked by the
super-index $I$ for the noise term. The It{\^o} version provides a
suitable base for the numerical integration of the Langevin equation
using a simple Euler scheme. The gradient term in \eq{langevin-ito} is
the combined result of using the It{\^o} version of the noise and a
term that compensates a spurious drift term arising from the no-slip
boundary conditions \cite{ermak:78,korn:07a}.

For our numerical simulations we discretize \eq{langevin-ito} in time
and use an Euler algorithm which is of first order in the time step
$\Delta t$
\begin{align}
  \label{langevin-euler}
  \Delta \boldv{X}_t = \boldv{F}\Delta t + k_B T\grad \mathsf{M}\Delta t +
  \boldv{g}(\Delta t) + \mathcal{O}(\Delta t^2)
\end{align}
where $\boldv{g}(\Delta t)$ has the same statistical properties as
$\boldv{g}_t^I$ from above.  In order to compute the
position-dependent mobility matrix $\mathsf{M}$ we use a scheme
presented by Jones et al.\ \cite{jones:92,jones:98} that provides
accurate results for all values of the height $z$. A detailed
description of the complete algorithm including the translational and
rotational update of the sphere can be found in Ref.~\cite{korn:07a}.

\subsection{Motor dynamics}
\label{sec:mdynamics}

\begin{figure}[t!]
  \begin{center}
    \resizebox{.65\linewidth}{!}{\includegraphics{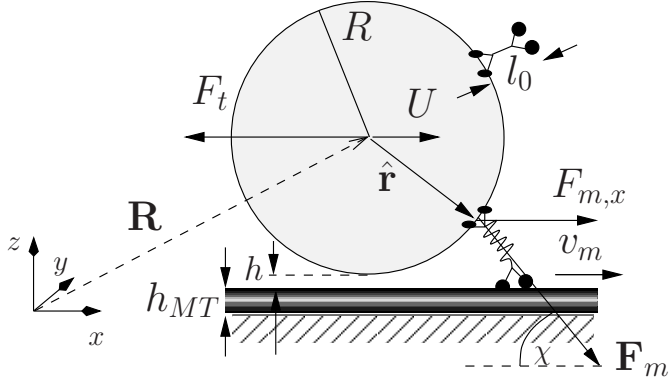}}
    \caption{A single sphere of radius $R$ and surface separation $h$
      from a planar wall.  The translational coordinates $\mathbf{R}$
      of the sphere are given relative to a reference frame that is
      fixed to the wall. The sphere is pulled by one molecular motor
      that is attached to the surface at position $\hat{\boldv{r}}$
      measured with respect to the center of the sphere. The bead is
      subject to the motor force $\boldv{F}_m$ with $x$-component
      $F_{m,x}$. In addition, an external force $F_t$ acts parallel
      to the filament, typically arising from an optical trap. The force
      unbalance between $F_{m,x}$ and $F_t$ leads to the bead
      velocity $U$.  The motor with resting length $l_0$ is firmly attached to
      the bead and can bind to and unbind from a MT and moves with velocity
      $v_m$. $\chi$ denotes the angle between motor and MT.
    \label{fig:setup}}
  \end{center}
\end{figure}
In our model the spherical cargo particle is uniformly covered with
$N_{tot}$ motor proteins. These molecules are firmly attached to the sphere
at their foot domains. Consequently, $N_{tot}$ is a constant in time. The
opposite ends of these molecules (their head domains) can reversibly attach to
the microtubule (MT), which is modeled as a line of equally spaced binding
sites for the motors covering the wall. A motor that is bound to the MT exerts a force and a
torque to the cargo particle. If $\boldv{r}_{h}$ and $\boldv{r}_{f}$ are the
positions of the head and foot domains of the motor, respectively, then the
force by the motor $\boldv{F}_m$ is given by:
\begin{align}
  \label{eq:singlemotorforce}
  \boldv{F}_m = \hat{\boldv{r}}_m F_m, \ha F_m=F(r_m), \ha \hat{\boldv{r}}_m = \frac{\boldv{r}_{h} -
    \boldv{r}_{f}}{\|\boldv{r}_{h} - \boldv{r}_{f}\|}, \ha r_m =  \|\boldv{r}_{h} - \boldv{r}_{f}\|,
\end{align}
where the absolute value $F_m$ of the motor force is given by the
force extension relation $F(x)$ for the motor protein. Throughout this
article we consider two variants of the force extension curve for the
polymeric tail of the motor. The harmonic spring model reads
\begin{align}
  \label{eq:spring}
  F(x) = \kk(x - l_0).
\end{align}
This means, a force is needed for both compression and extension of
the motor protein.  Actually, it was found that kinesin exhibits a
non-linear force extension relation \cite{svoboda:93}, with the spring
constant varying between $\kk = 0.2\cdot 10^{-4}$~N/m for small
extensions and $\kk = 0.6 \cdot 10^{-4}$~N/m for larger extensions
\cite{svoboda:93,gibbons:01}. For extensions close to the contour
length the molecule becomes infinitely stiff \cite{duke:96}. For small
extensions, however, the harmonic approximation works well. Alternatively,
we consider the cable model
\begin{align}
  \label{eq:cable}
  F(x) = \kk (x - l_0) \Theta(x-l_0),\ha  \Theta(x) = \left \{
  \begin{array}{l}1,\ha x > 0\\0,\ha \mbox{else} \end{array}
  \right..
\end{align}
In the cable model force is only built up when the motor is extended
above its resting length $l_0$. In the compression mode, i.\,e. when
the actual motor length is less then the resting length, no force
exists. The cable model can be seen as the simplest model for a
flexible polymer.

Besides the force each motor attached to the MT also exerts a torque on the
cargo particle.  With $\hat{\boldv{r}}$ being the position of the motor
foot relative to the center of the sphere (cf. \fig{fig:setup}), this torque
reads
\begin{align}
  \label{eq:singlemotortorque}
  \boldv{T}_m = \hat{\boldv{r}} \times \boldv{F}_m.
\end{align}
The combined force/torque vector $\boldv{F} = (\boldv{F}_m,\boldv{T}_m)^T$
enters the Langevin algorithm \eq{langevin-euler}. 

In addition to the firm attachment of the motors to the sphere, each
motor can in principle reversibly bind and unbind to the MT. We model
these processes as simple Poissonian rate processes in similar a
fashion as it is done for modeling of formation and rupture of
receptor ligand bonds in cell adhesion (e.\,g., \cite{hammer:92}). The
motor head is allowed to rotate freely about its point of fixation on
the cargo. The head is therefore located on a spherical shell with
radius given by the motor resting length $l_0$.  However, in contrast
to the anchorage point on the cargo particle the head position of the
motor is not explicitely resolved by the algorithm. In order to model
the binding process we introduce a capture length $r_0$. A motor head
can then bind to the MT with binding rate $\pi_{ad}$ whenever the
spherical shell of radius $l_0$ and thickness $r_0$ around the motor's
anchorage point has some overlap with a non-occupied binding site on
the MT.  The MT's binding sites are identified by the tubulin building
block of length $\delta = 8$~nm, so we choose $r_0 = \delta/2$. Note
that binding rate and binding range are complementary quantities and that
a more detailed modeling of the binding process would have to yield
appropriate values for both quantities. If
the overlap criterion is fulfilled within a time interval $\Delta t$,
the probability for binding $p_{on}$ within this time step is $p_{on}
= 1 - \exp(-\pi_{ad} \Delta t)$. With a standard \emph{Monte-Carlo
  technique} it is then decided whether binding occurs or not: a
random number $rand$ is drawn from the uniform distribution on the
interval $(0,1)$ and in the case $p_{on} > rand$ binding occurs. If
the overlap criterion is fulfilled for several binding sites, then
using the Monte-Carlo technique it is first decided whether binding
occurs and then one of the possible binding sites is randomly chosen.

\begin{table}[t!]
  \begin{center}
    \begin{tabular}{lccc}
      \hline\hline
      Parameter& typical value&meaning &reference \\
      \hline
      $R$&$1~\mu$m&bead radius&\\
      $\epsilon_{0}$&$1$~s$^{-1}$&unstressed escape rate&\cite{klumpp:05}\\
      $\pi_{ad} $&$\mathbf{5}$~s$^{-1}$&binding rate&\cite{c:ledu04, klumpp:05}\\
      $F_d$&3~pN&detachment force&\cite{klumpp:05,block:00}\\
      $\kk$&$10^{-5}\ldots\mathbf{10^{-4}}\ldots 10^{-3}$~N/m &motor spring constant&
      \cite{rohrbach:01,duke:96,gibbons:01}\\
      $\delta$&8~nm&kinesin step length&\cite{coy:99}\\
      $v_0$&$1~\mu$m/s&maximum motor velocity&\cite{klumpp:05}\\
      $\lambda_s^0 := v_0/\delta$&125~s$^{-1}$&forward step rate&\\
      $r_0 $&$\delta/2$&capture radius&\\
      $F_s$&$5\ldots\mathbf{6}\ldots8$~pN&stall force&\cite{block:99}\\
      $l_0$&50,65,80~nm& (resting) length &\cite{diez:06,schliwa:03}\\
      $h_{MT}$&24~nm&microtubule diameter&\cite{limberis:01}\\
      \hline\hline
    \end{tabular}  
    \caption{Parameters used for adhesive motor dynamics. 
      For ambient temperature we use $T = 293$~K, for viscosity $\eta = 1$~mPa\,s
      (if not otherwise stated). If a range is given, then figure in bold face
      denotes the value used in the numerical simulations.}
    \label{tab:parameter}
  \end{center}
\end{table}
Each motor bound to the MT can unbind with escape rate
$\epsilon$. In single motor experiments it was found that the escape rate increases
with increasing motor force \cite{block:00}. This force
dependence can be described by the Bell equation \cite{bell:78,klumpp:05}
\begin{align}
  \label{eq:bell}
  \epsilon = \epsilon_0 \exp(F_m/F_d),
\end{align}
with the unstressed escape rate $\epsilon_0$ and the detachment force $F_d$.
Because the details of forced motor unbinding from a filament are not
known, here we make the simple assumption that the unbinding pathway
is oriented in the direction of the tether.
We set $\epsilon = \epsilon_0$ whenever the motor is compressed.

The major conceptual difference between a motor connecting a sphere
with a MT and a receptor-ligand bond is that a motor can actively step
forward from one binding site to the next with step length $\delta$
(the length of the tubulin unit). The mean velocity $v_0$ of an
unloaded kinesin motor is about $v_0 = 1~\mu$m/s depending on ATP
concentration \cite{block:99}. If the motor protein is mechanically
loaded with force opposing the walking direction, the motor velocity
$v_m$ is decreased. For a single kinesin  molecule that is attached to
a bead on which a trap force $F_t$ pulls, the velocity was found
experimentally to decrease approximately linearly \cite{svoboda:94a,
  block:99}:
\begin{align}
  \label{eq:trapforcerelation}
  v_m = v_0\left(1 - \frac{F_t}{F_s}\right), \ha 0 < F_t < F_s,
\end{align} 
with the stall force $F_s$ and the trap force $F_t$ acting
antiparallel to the walking direction.
Different experiments have reported stall forces between 5 and 8 pN.
Changes in this range are not essential for our results and therefore
we use the intermediate value $F_b = 6\ pN$. If the force is higher than the
stall force, kinesin motors walk backwards with a very low velocity
\cite{Carter_Cross2005}, which we will neglect in the following.
Finally, for assisting forces, i.e. if the motor is pulled forward,
the effect of force is relatively small
\cite{block:04,Carter_Cross2005}. In order to derive an expression
similar to \eq{eq:trapforcerelation} for our model, we have to
identify the proper term that replaces $F_t$ in
\eq{eq:trapforcerelation}. First, we rewrite \eq{eq:trapforcerelation}
as $v_m = \mu_m (F_s - F_t)$,\label{page:mobmotor2} with some
\emph{internal} motor mobility coefficient $\mu_m := v_0/F_s$. This
version of \eq{eq:trapforcerelation} allows us to interpret the motor
head as an over-damped (Stokesian) particle that constantly pulls with
the force $F_s$ against some external load $F_t$ resulting in
the effective velocity $v_m$.  According to \eq{eq:singlemotorforce} a
motor pulls with force $\boldv{F}_m$ on the bead, so we can identify
the ``load'' to be $-F_{m,x}$, where $F_{m,x}$ is the $x$-component of
$\boldv{F}_m$ and the minus sign accounts for Newton's third law
(\emph{actio = reactio}). Thus, we obtain the following piecewise
linear force velocity relation for the single motor (see also
\fig{fig:model}):
\begin{figure}[t!]
  \begin{center}
    \resizebox{.46\linewidth}{!}{\includegraphics{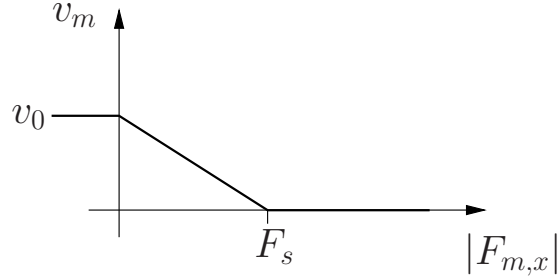}}
    \caption{Force velocity relation for a single motor:
      velocity $v_m$ as a function of load $|F_{m,x}|$ according to
      \eq{force-velocity} with maximum velocity $v_0$ and stall force $F_s$.
      \label{fig:model}}
  \end{center}
\end{figure}
\begin{align}
  \label{force-velocity}
  v_m = v_0 \left\{ \begin{array}{ccc}  
    1 &\;\;\;\mbox{if}\;\;\; &\boldv{F}_m\cdot\boldv{e}_x \leq 0\\ 
    1 - \frac{|F_{m,x}|}{F_s} &\;\;\;\mbox{if}\;\;\;
    &  0 < \boldv{F}_m\cdot\boldv{e}_x < F_s,\\ 
    0 &\;\;\;\mbox{if}\;\;\;&\boldv{F}_m\cdot\boldv{e}_x \geq F_s
  \end{array}\right .
\end{align}     
where $\boldv{e}_x$ is the walking direction of the motor, see
\fig{fig:setup}.  Thus, if the motor pulls antiparallel to its walking
direction on the bead, it walks with its maximum speed $v_0$. If it is loaded
with force exceeding the stall force $F_s$, it stops. For intermediate
loadings the velocity decreases linearly with load force. \eq{force-velocity}
defines the mean velocity of a motor in the presence of loading force.
We note that our algorithm also allows us to implement more complicated force-velocity
relations and is not restricted to the piecewise linear force-velocity relation.
It is used here because we do not focus on a specific kinesin motor
and because it is easy to implement in the computer simulations.
In practice the motor walks with discrete steps of length $\delta$. In the
algorithm we account for this by defining a step rate $\lambda_s :=
v_m/\delta$. The decision for a step during a time interval $\Delta t$ is then
made with the same Monte-Carlo technique used to model the binding and
unbinding process of the motor head. A step is rejected if the next binding
site is already occupied by another motor (mutual exclusion)
\cite{lipowsky:01}.

\subsection{Bead versus motor friction}
\label{sec:friction}

In principle, the velocity of a single motor $v_m$ pulling the sphere and the
component in walking direction of the sphere's velocity $U$ (see
\fig{fig:setup}) are not the same. For an external force $F_t$ (against
walking direction) in the pN range acting on the sphere and $F_{m,x}$ being
the motor force in walking direction, the bead velocity $U$ is given by $U =
\mu_{xx}^{tt}(F_{m,x} - F_t)$,\label{page:mobmotor}. Here, $\mu_{xx}^{tt}$
denotes the corresponding component of the mobility matrix $\mathsf{M}$ of the
sphere (cf. \eq{langevin-ito}) evaluated at the height of the sphere's center
with super indices $tt$ referring to the translational sector of the matrix and
sub indices $xx$ referring to the responses in $x$-direction. On the other
hand from \eq{force-velocity} it follows that the motor head moves with
velocity $v_m = \mu_m(F_s - F_{m,x})$. Only in the stationary state of motion
the two speeds are equal, $U = v_m$, and we obtain the force with which the
motor pulls (in walking direction) on the bead:
\begin{align}
  \label{eq:motorforce}
  F_{m,x} = \frac{\mu^{tt}_{xx}}{\mu_m + \mu^{tt}_{xx}}F_t 
  + \frac{\mu_m}{\mu_m + \mu^{tt}_{xx}}F_s.
\end{align}
Thus, if the internal friction of the motor is large
compared to the viscous friction of the sphere, i.\,e., $1/\mu_m \gg
1/\mu^{tt}_{xx}$, the second term in \eq{eq:motorforce} can be
neglected and one has $F_{m,x} \approx F_t$. That means, only the trap
force pulls on the motor. If $\mu_m \approx \mu^{tt}_{xx}$ both terms
in \eq{eq:motorforce} are of the same order of magnitude. Then, both
external load $F_t$ and the friction force on the bead will influence
the motor velocity. Experimentally, these prediction can be checked in
bead assays by varying the viscosities of the medium (e.\,g., by
adding sugar like dextran or Ficoll \cite{springer:01}). Numerically,
we can vary $\eta$ in the adhesive motor dynamics algorithm.

When several motors are simultaneously pulling, they can cooperate by sharing
the load. Assuming the case that $n$ motors are attached
to the MT which \emph{equally} share the total load, then the force in
$x$-direction exerted on the bead is $n F_{m,x}(n)$ with $F_{m,x}$ being again
the individual motor force. In the stationary state with $U = v_m$ we have 
(with external load $F_t = 0$):
\begin{align}
  F_{m,x}(n) = \frac{\mu_m}{n\mu_{xx}^{tt} + \mu_m} F_s.
\end{align}
The mean bead velocity $U(n)$ with $n$ equally pulling motors is $U(n)
= n\mu_{xx}^{tt} F_{m,x}(n)$.  Thus, in general, the velocity of the
beads will increase with increasing number $n$ of pulling motors.
However under typical experimental conditions in vitro, where bead
movements are probed in aqueous solutions, the internal motor friction
$1/\mu_m$ dominates over the viscous friction of the bead,
$1/\mu_{xx}^{tt}$, and the velocity becomes independent of the number
of motors. Only if the bead friction becomes comparable to the
internal motor friction, the velocity exhibits an appreciable
dependence on the motor number. This dependence can be illustrated by
considering the ratio of $U(n)$ and $U(1)$:
\begin{align}
  \label{eq:velratio}
  \frac{U(n)}{U(1)} = \frac{n \mu_{xx}^{tt} + n \mu_m}
       {n \mu_{xx}^{tt} + \mu_m} = \frac{n/\mu_m + n/\mu_{xx}^{tt}}
       {n/\mu_m + 1/\mu_{xx}^{tt}}.
\end{align} 
In the opposite limit, $n/\mu_m \ll 1/\mu_{xx}^{tt}$, i.\,e., when the
viscous bead friction is very large and dominates over the internal
motor friction, \eq{eq:velratio}, leads to $U(n) \approx n U(1)$, and
the bead velocity increases linearly with $n$ \cite{klumpp:05}.

\subsection{Vertical forces}
\label{sec:vertical}

We note that, although we are mainly interested in the $x$-component
of the motor force $\boldv{F}_m$, i.e.\ the component parallel to the
microtubule, which enters the force--velocity relation, the motor
force also has a component perpendicular to the microtubule, see
\fig{fig:setup}. This force component tends to pull the bead towards
the microtubule and thus to the surface, whenever the bead is
connected to the microtubules by a motor. This force is balanced by the microtubule repelling
the bead. Additionally, if the bead touches the filament or the wall,
diffusion can only move the bead away from the wall. In case of the full spring model, compressed motors can
also contribute a repulsion of the bead from the wall. If viscous friction is
strong, the normal component of the force arises mainly from the
microtubule. In that case, the distance $h$ between the bead and the
microtubule is very small, $\mean h \approx 0$. For small viscous friction,
thermal fluctuations play a major role and lead to non-zero distances
between the bead and the microtubule, as discussed further in
\sec{sec:walking}.

All parameters used for the adhesive motor dynamics simulations together with
typical values are summarized in \tab{tab:parameter}. For the numerical
simulations we non-dimensionalize all quantities using $R$ for the length
scale, $1/\epsilon_0$ for the time scale and the detachment force $F_d$ as
force scale. 

\subsection{Master equation approach}
\label{sec:theory}

When no external load is applied a motor walks on average a time
$1/\epsilon_0$ before it detaches and the cargo particle might diffuse
away from the MT. When several motors on the cargo can bind to the MT
the mean run length dramatically increases as was previously
shown with a master equation approach \cite{klumpp:05}. For the sake
of later comparison to our simulations we briefly summarize some of
these results in the following.

Let $P_i$ be the probability that $i$ motors are simultaneously bound to the
MT with $i = 0,\ldots,\Nb$ and $\Nb$ being the maximum number of motors that
can bind to the MT simultaneously. Assuming the system of $\Nb$ motors is in a
stationary state and the total probability of having $i = 0,\ldots,\Nb$ motors
bound is conserved then the probability flux from one state to a neighboring
state is zero. This means that the probability $P_i$ of having $i$ bound
motors can be calculated by equating forward and reverse fluxes
\begin{align}
  \label{eq:steadystate}
  (\Nb - i)\pi_{ad} P_i = (i+1)\epsilon P_{i+1},\ha i = 0,\ldots,\Nb-1,
\end{align}
where it is assumed that the escape rate $\epsilon$ is a constant with respect
to time. The solution to \eq{eq:steadystate} is given by
\begin{align}
 P_i = \left(\begin{array}{c} \Nb\\ i\end{array}\right)
   \frac{\epsilon^{\Nb-i}\pi^i}{(\epsilon + \pi)^{\Nb}},\ha i = 0,\ldots,\Nb. 
\end{align}
The probability that $i$ motors are simultaneously pulling under the condition
that at least one motor is pulling is $P_i/(1 - P_0)$ for $i = 1,\ldots,\Nb$.
Then, the mean number of bound motors $\mNb$ (given that at least one motor is
bound) is \cite[Eq.~{[13]}]{klumpp:05}
\begin{align}
  \label{mean_motor}
  \mNb = \sum_{i = 1}^{\Nb} \frac{iP_i}{1 - P_0} = 
  \frac{(\pi_{ad}/\epsilon)\left[1 +
  \pi_{ad}/\epsilon\right]^{\Nb-1}}{\left[1 + \pi_{ad}/\epsilon\right]^{\Nb} - 1}\Nb.
\end{align}
The effective unbinding rate $\epsilon_{eff}$, i.\,e., the rate with which the
system reaches the unbound state, is determined from $\epsilon_{eff} (1 - P_0)
= \pi_{ad} P_0$. This quantity can also be identified with the inverse of the
mean first passage time for reaching the unbound state, when starting with one
motor bound \cite{klumpp:05}. If the medium viscosity is small, i.\,e.,
similar to that of water, and no external force is pulling on the bead, we
assume that the velocity of the bead $U$ does not depend on the number of
pulling motors. The mean run length, that is the mean distance the cargo
is transported by the motors in the case that initially one motor was bound,
is then the product of mean velocity $U$ and mean lifetime
($1/\epsilon_{eff}$) \cite[Eq.~{[14]}]{klumpp:05}:
\begin{align}
  \label{mean_walk}
  \dxb = \frac{U}{\epsilon_{eff}} = \frac{U}{\Nb\pi_{ad}}\left[\left(1 +
  \frac{\pi_{ad}}{\epsilon}\right)^{\Nb} - 1\right].
\end{align}
For kinesin-like motors at small external load with $\pi_{ad} \gg \epsilon$
this expression can be approximated by $\dxb \approx (U/\Nb
\epsilon)(\pi_{ad}/\epsilon)^{\Nb-1}$, i.\,e., the mean run length grows
essentially exponentially with $\Nb$.  In the stationary state the bead
velocity $U$ and the motor velocity $v_m$ are equal.  For no external load and
small viscous friction on the bead one can approximate $\epsilon \approx
\epsilon_0$ in \eq{mean_motor} and \eq{mean_walk}.

\subsection{Mean run length for a spherical cargo particle}
\label{sec:poisson}

\begin{figure}[t!]
  \begin{center}
    \resizebox{.43\linewidth}{!}{\includegraphics[angle=0]{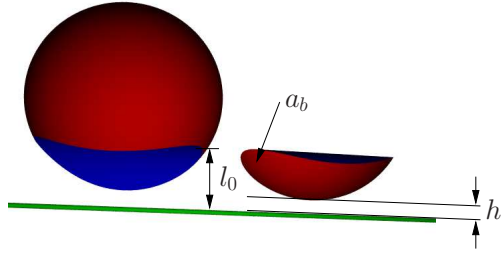}}
    \caption{Illustration of the area fraction $a_b$ of the sphere
      (cut and placed besides the sphere) on which
      motor proteins can reach the MT (thin cylinder). $a_b$ depends
      in a geometrical fashion on the minimum distance $h$ between the sphere and the MT and on
      the resting length $l_0$.
      \label{fig:area}}
  \end{center}
\end{figure}
In contrast to the master equation model in which one fixes the
maximal number of bound motors $\Nb$, in the computer simulations only
the total number $N_{tot}$ of motors on the spherical cargo particle
is fixed. A similar situation arises in experiments where only the
total amount of molecular motors on the sphere is measured
\cite{beeg:08} (but not in experiments with defined multi-motor
complexes \cite{rogers09}). If $A_b$ is the area on the sphere's
surface that includes all points being less than $l_0$ apart from the
MT (cf. \fig{fig:area}), we expect on average $n_b = N_{tot}a_b$
motors to be close enough to the MT for binding, with the reduced area
$a_b := A_b/(4\pi R^2)$. While the master equation model assumes a
fixed $\Nb$, in the simulations the maximum number of motors that can
simultaneously bind to the MT is a fluctuating quantity about the mean
value $n_b$. In the simulations the motors are uniformly distributed
on the cargo, thus the probability distribution function $P(k)$ for
placing $k$ motors inside the above defined area fraction $a_b$ is a
binomial distribution. As we have $l_0 \ll R$, $a_b$ is small and
$P(k)$ is well approximated by the \emph{Poissonian} probability
distribution function
\begin{align}
  \label{poisson}
  P(k, n_b) = \frac{n_b^k}{k!} \exp(-n_b),\hspace{2cm} 
  n_b = N_{tot}a_b.
\end{align} 
Thus, given a fixed number $N_{tot}$ of motors on the sphere the number of
motors that are initially in binding range to the MT might be different from
run to run. In addition, because of thermal fluctuations and torques that the
motors may exert on the cargo particle the relative orientation of the sphere
changes during a simulation run. With the change of orientation also the
number of motors in binding range to the MT is not a constant quantity during
one run.

In order to make simulation results for the mean run length and the mean
number of bound motors comparable with the theoretical predictions
\eq{mean_walk} and \eq{mean_motor}, respectively, we have to average over
different $\Nb$. Neglecting fluctuations in the number of motors that are in
binding range to the MT during one simulation run we perform the average with
respect to the Poisson distribution, \eq{poisson}. Averaging the mean walking
distance $\dxb(\Nb)$ from \eq{mean_walk} over all $\Nb$ with weighting factors
given by \eq{poisson} we obtain the following expression ($ n_b = a_b
N_{tot}$):
\begin{align}
  \label{meanmeanwalk}
  \left . \mean{\dxb}_{Poisson} = 
  \sum\limits_{\Nb = 1}^{\Nbmax}\frac{n_b^{\Nb -1}}{(\Nb - 1)!}   
  \frac{U}{\Nb\pi_{ad}}\left[\left(1 + \frac{\pi_{ad}}{\epsilon_0}\right)^{\Nb} - 1\right]
  \right/\sum\limits_{\Nb = 1}^{\Nbmax}\frac{n_b^{\Nb -1}}{(\Nb - 1)!}.
\end{align}
We note that during the initialization of each simulation run we place
one motor on the lower apex of the sphere and then distribute the
other $N_{tot} - 1$ motors uniformly over the whole sphere (see
appendix \ref{appendix:distribution} for a detailed description of the
procedure).  For this reason $P(\Nb - 1,n_b)$ denotes the probability
of having in total $\Nb$ motors inside the area fraction $a_b$.
Furthermore, we introduced a cutoff $\Nbmax$ of maximal possible
motors ($N_{tot}$ is obviously an upper limit for $\Nbmax$).  In the
limit $\Nbmax \rightarrow \infty$ we have $\mean{\dxb}_{Poisson} =
U(e^{\pi_{ad} n_b/\epsilon_0} - 1)/(\pi_{ad} n_b)$.

In a similar way we can calculate the Poisson-averaged mean number of bound
motors $\mean{\mNb}_{Poisson}$. For this it is important to include the
correct weighting factor \cite{korn:phd}. From an ensemble of many simulation
runs those with larger run length contribute more than those with
smaller run length. For $\N$ simulation runs the mean number of bound
motors is obtained as $\mean{\mNb}_{sim} = \sum_i t_i n(i)/\sum_i t_i$, where
$t_i$ is a period of time during which $n(i)$ motors are bound and the sum is
over all such periods of time. Assuming the bead velocity $U$ to be a
constant, the time periods $t_i$ can also be replaced by the run lengths
$\Delta x_{b,i}$ during $t_i$. Picking out all simulation runs with a fixed
$\Nb$, their contribution to the sum is the mean number of bound motors $\mNb$
times the total run length of beads with given $\Nb$. The latter is the
mean run length $\dxb_{\Nb}$ times the number of simulation runs with
the given $\Nb$ (for sufficiently large $\N$). Clearly, the fraction of runs
with given $\Nb$ is the probability $P(\Nb - 1, n_b)$ introduced in
\eq{poisson}.  Consequently, we obtain again with a truncation of the sum at
some $\Nbmax \leq N_{tot}$
\begin{align}
  \label{meanmeanbound}
  \mean{\mNb}_{Poisson} = \sum_{\Nb = 1}^{\Nbmax} \bar P(\Nb) \mNb(\Nb)=
  \frac{\sum_{\Nb = 1}^{\Nbmax} P(\Nb-1, n_b) \dxb_{\Nb} \mNb(\Nb)}
       {\sum_{\Nb = 1}^{\Nbmax} P(\Nb-1, n_b) \dxb_{\Nb}}.
\end{align}
Here we introduced the probability $\bar P(n)$ for having $n$ motors in
binding range to the MT when picking out some cargo particle from a large
ensemble of spheres. Explicitely, this probability is given by 
\begin{align}
  \label{eq:probn}
  \bar P(n) = 
  \frac{\frac{U}{\pi_{ad}}((1+\pi_{ad}/\epsilon)^n - 1)\frac{n_b^{n-1}}{n!}e^{-n_b}}
       {\sum_{i = 1}^{\Nbmax}\frac{U}{\pi_{ad}}((1+\pi_{ad}/\epsilon)^i -
  1)  \frac{n_b^{i-1}}{i!}e^{-n_b}} 
  = \frac{((1+\pi_{ad}/\epsilon_0)^n -1)n_b^{n-1}/n!}
  {\sum_{i = 1}^{\Nbmax}((1+\pi_{ad}/\epsilon_0)^i - 1)n_b^{i-1}/i!}.
\end{align}

\subsection{Computer simulations}
\label{sec:comp_simulations}

We use the following procedure for the computer simulations.  In each
simulation run the sphere is covered with $N_{tot}$ motors. Initially,
one motor, located at the lowest point of the sphere, is attached to
the microtubule such that the distance of closest approach $h$ between
the sphere and the microtubule is given by the resting length of the
motor, i.\,e., $h = l_0$. The other $(N_{tot} - 1)$ motors are
uniformly distributed on the sphere's surface (cf.
\sec{appendix:distribution}).  When the motor starts walking, it pulls
the sphere closer to the MT because there is a $z$-component in the
force exerted on the sphere by the motor stalk (which is strained
after the first step). Then, other motors can bind to the MT. The
system needs some time to reach a stationary state of motion, so
initially the motor velocity $v_m$ and the bead velocity $U$ are not
the same (for reasons of comparison a fixed initial position is
necessary; other initial positions have been tested but initialization
effects were always visible). In principle a simulation run lasts
until no motor is bound. For computational reasons, each run is
stopped after $2\cdot 10^4$~s (which is rarely reached for the
parameters under consideration).  For each run quantities like the
mean number of bound motors $\mNb$, the walking distance $\Delta x_b$
and the mean distance of closest approach $\mean{h}$ between sphere
and MT are recorded.

\section{Results}
\label{sec:results}

\subsection{Single motor simulations}
\label{sec:single}

\begin{figure}[t!]
  \begin{center}
    \begin{tabular}{c@{\hspace{.04\linewidth}}c}
    \resizebox{.46\linewidth}{!}{\includegraphics[angle=0]{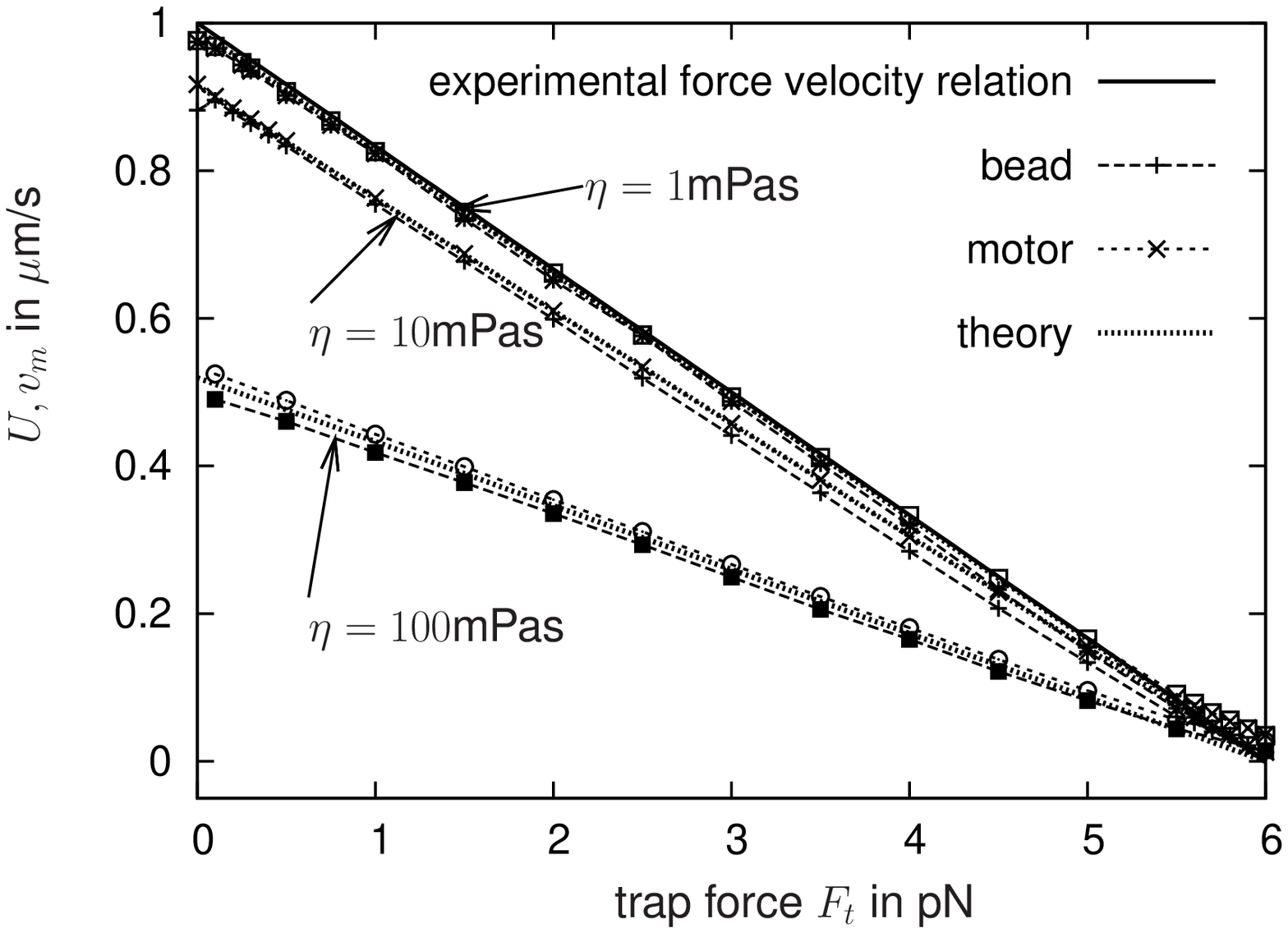}}&
    \resizebox{.46\linewidth}{!}{\includegraphics[angle=0]{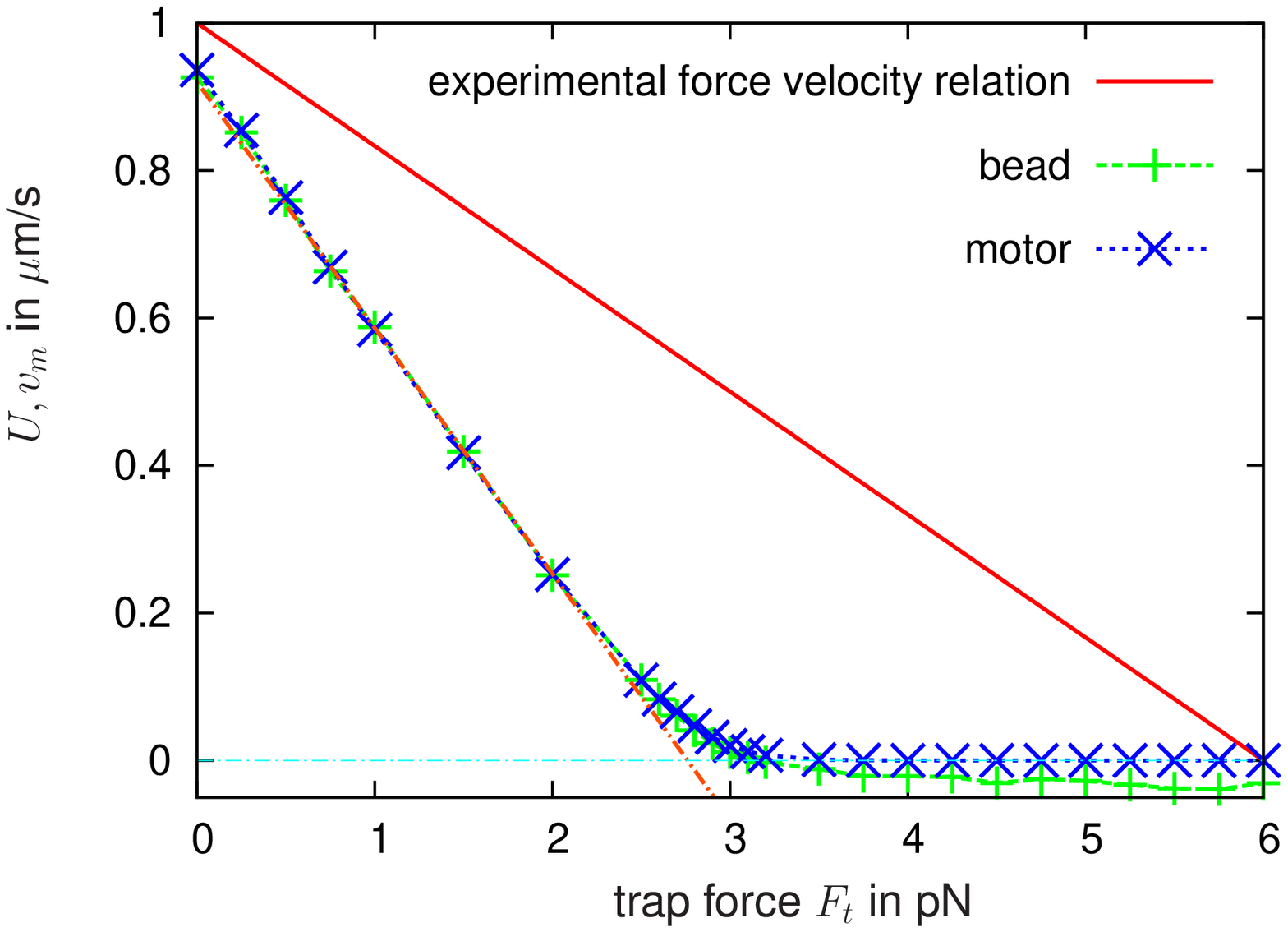}}\\ (a) &
    (b) \\
    \end{tabular}
    \caption{(a) Measured force velocity relation of a single motor
      (with $l_0 = 80$~nm) pulling a sphere of radius $R = 1~\mu$m for
      three different viscosities $\eta = 1,10,100$~mPa\,s.  Shown is the
      relation according to
      \eq{eq:trapforcerelation}, the actual measured force
      velocity relation of the motor head and the bead center,
      respectively, and the theoretical prediction according to
      \eq{force-velocity} and \eq{eq:motorforce}.  (b) The
      measured force velocity relation for $\eta = 1$~mPa\,s is shown
      where in \eq{force-velocity} not $|F_{m,x}|$ but
      $\|\boldv{F}_{m}\|$ is used. The dotted line emphasizes the
      linear decrease of the velocity. The negative velocity of the
      bead at large $F_t$ results from thermal fluctuations.
      Fluctuations against walking direction increase the escape
      probability. In case of escape they cannot be compensated by
      fluctuations in walking direction.  (Numerical parameters:
      $\Delta t = 10^{-5}$, number of runs $\N = 2\cdot 10^3 - 9\cdot
      10^4$.)
      \label{res:forcevelsingle}}
  \end{center}
\end{figure}

In \sec{sec:mdynamics} we defined a force-velocity relation for the
single motor. In this section we perform simulation runs with a single
motor, i.\,e., $N_{tot} = 1$, and measure the effective force velocity
relation by tracking the position of the sphere.  Inserting
\eq{eq:motorforce} into \eq{force-velocity} provides a prediction for
the velocity of a bead subject to one pulling motor and an external
trap force $F_t$.  In \fig{res:forcevelsingle}a this prediction is shown
for three different values of viscosity together with the actual
measured velocities during the simulations. More precisely, we
measured the mean velocity of the bead and the motor obtained from a
large number of simulation runs (to avoid effects resulting from the
initial conditions we first allowed the relative position/orientation
of bead and motor to ``equilibrate'' before starting the actual
measurement). The mean velocity is then given as the total (summed up
over all simulation runs) run length divided by the total walking
time. The good agreement between the numerical results and the
theoretical predictions provides a favorable test to the algorithm. At
$\eta = 1$~mPa\,s (the viscosity of water), friction of the bead has
almost no influence on the walking speed. At hundred times larger
viscosities, however, bead friction reduces the motor speed to almost
half of its maximum value already at zero external load. Although the
velocities of the motor and the bead are expected to be equal,
\fig{res:forcevelsingle}a shows that the motor is slightly faster than
the bead.  This is a result of the discrete steps of the motor and can
be considered as a numerical artefact: at the moment the motor steps
forward the motor stalk is slightly more stretched (loaded) than
before the step, therefore, the escape probability is increased.  The
result of unbinding at the next time step would then be that the bead
moved a distance $\delta$ less than the motor. For loads close to the
stall force the observed velocity is slightly larger than the
prediction, which is a result of thermal fluctuations of the bead in
combination with the stepwise linearity of the force velocity
relation: a fluctuation in walking direction slightly reduces the load
on the motor, thus increasing the step rate, whereas fluctuations
against walking direction lead to zero step rate.

It was observed by Block et al. that vertical forces on the bead (i.\,e., in
$z$-direction) also reduce the velocity of the motor \cite{block:04}. But the
same force that leads to stall when applied antiparallel to the walking
direction has a rather weak effect on the motor velocity when applied in
$z$-direction. Using the absolute value of the total force of the motor
$\|\boldv{F_m}\|$ in \eq{force-velocity} instead of its $x$-component
$|F_{m,x}|$, we measure a force velocity relation as shown in
\fig{res:forcevelsingle}b.  Again, the velocity decreases essentially linearly
with applied external force, but stalls already at around $F_t \approx F_s/2$
because of vertical contributions of the force $\|\boldv{F}_m\|$.  As the
effect of vertical loading reported in Ref.~\cite{block:04} seems to be much
weaker than that shown in \fig{res:forcevelsingle}b, we reject this choice of
force velocity relation.

\begin{figure}[t!]
  \begin{center}
    \resizebox{.46\linewidth}{!}{
      \includegraphics[width=\linewidth]{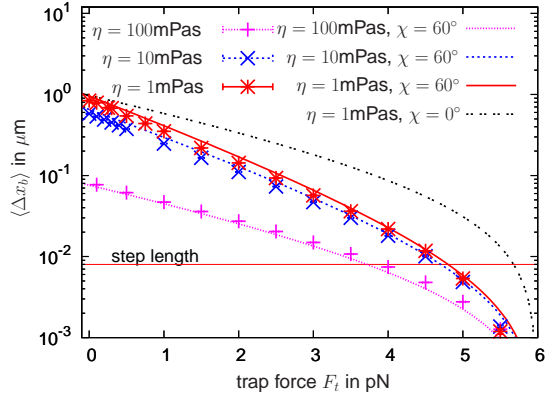}}
    \caption{Mean run length $\dxb$ of a bead pulled by a single motor
      as a function of an external force on the bead $F_t$ and for three
      different viscosities $\eta = 1,10,100$~mPa\,s. The lines give the
      theoretical predictions according to \eq{eq:singlemwd} assuming an
      angle of $60^\circ$ between the motor and the MT.  For comparison also
      the theoretically predicted $\dxb$-curve for $\chi = 0$ is shown (double
      dotted line).\label{res:meanwalkingsingle}}
  \end{center}
\end{figure}
From the simulations carried out for \fig{res:forcevelsingle}a we can also
obtain the mean run length $\mean{\Delta x_b}$ for a single motor as a
function of external load.  The results are shown in
\fig{res:meanwalkingsingle}.  Using \eq{force-velocity} and the Bell equation,
\eq{eq:bell}, we obtain
\begin{align}
  \label{eq:singlemwd}
  \dxb = \frac{v_m}{\epsilon} = 
  \frac{v_0}{\epsilon_0}\frac{ 1 - |F_{m,x}|/F_s}{\exp(\|\boldv{F}_m\|/F_d)}.
\end{align}
The numerical results shown in \fig{res:meanwalkingsingle} fit very well to the
theoretical prediction of \eq{eq:singlemwd} when assuming the angle
$\chi$ between the motor and the MT to be $\chi = 60^\circ$. The angle $\chi$
depends on the bead radius $R$, the resting length $l_0$ \cite{block:03} and
the polymer characteristics of the motor protein, e.\,g., its stiffness $\kk$.

\subsection{Run length for several motors pulling}
\label{sec:walking}

\begin{figure}[t!]
  \begin{center}
    \begin{tabular}{c@{\hspace{.04\linewidth}}c}
    \resizebox{.46\linewidth}{!}{\includegraphics[angle=0]{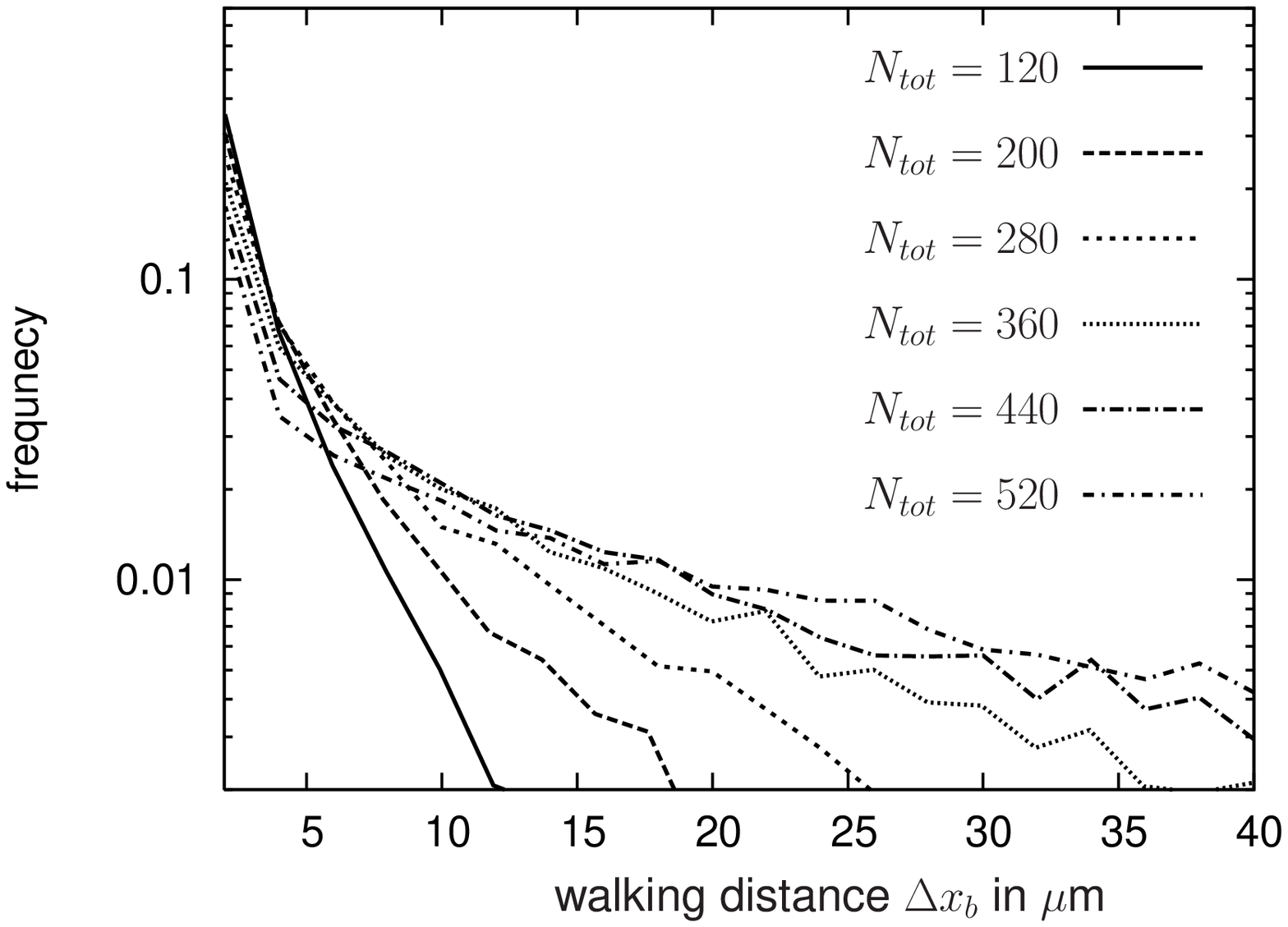}}&
    \resizebox{.46\linewidth}{!}{\includegraphics[angle=0]{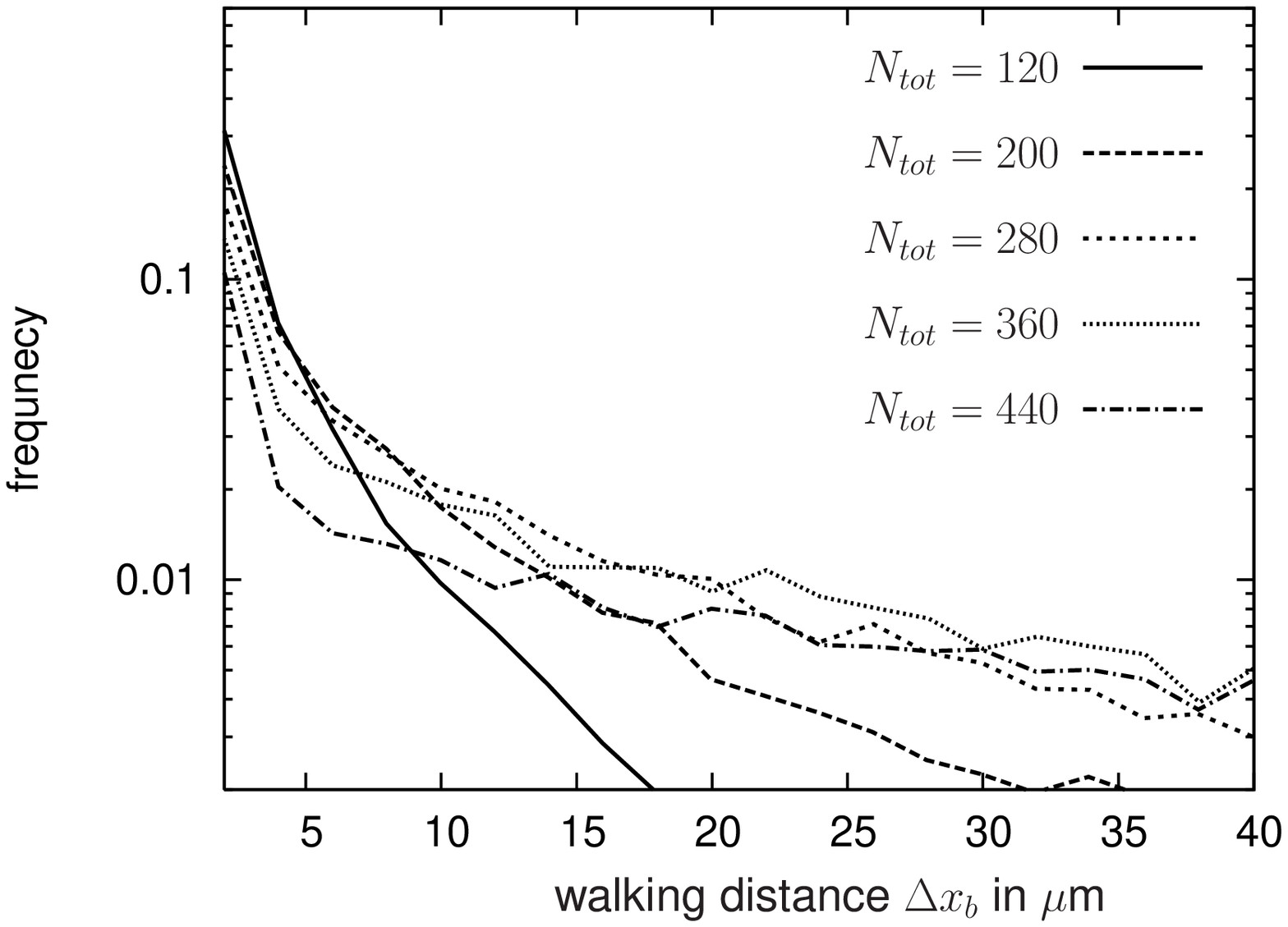}}\\ 
    (a) & (b) \\
    \end{tabular}
    \caption{Distribution of run lengths $\Delta x_b$ in
      semi-logarithmic scale for different values of motor coverage $N_{tot}$.
      The motor protein is modeled as a harmonic
      spring according to \eq{eq:spring}.  (a) Resting length of
      the motor protein $l_0 = 50$~nm.  (b) Resting length of the
      motor protein $l_0 = 65$~nm.  (Numerical parameters: time step
      $\Delta t = 10^{-5}/\epsilon_0$, number of simulation runs $\N\approx 10^4$.)
      \label{res:histos}}
  \end{center}
\end{figure}
We now measure the run length distributions and
the mean run length as a function of motor coverage $N_{tot}$. For motors
modeled as springs according to \eq{eq:spring} with two different resting
lengths $l_0 = 50,65$~nm the run length distributions are shown in
\fig{res:histos}a,b. For each value of $N_{tot}$ the run length was
measured about $\N = 10^4$ times. The simulations turn out to be very costly,
especially for large $N_{tot}$ as the mean run length increases
essentially exponentially with the number of pulling motors
(cf. \eq{mean_walk}). From \fig{res:histos} we see that the larger $N_{tot}$,
the more probable large run lengths are, resulting in distribution
functions that exhibit a flatter and flatter tail upon increasing $N_{tot}$.
\fig{res:histos} nicely shows that the shape of the distributions depends not
only on the total number of motors $N_{tot}$ attached to the sphere but also
on the resting lengths. Given the same $N_{tot}$ we can see that longer
run lengths are more probable for the longer resting length $l_0 =
65$~nm shown in \fig{res:histos}b. This can simply be explained by the fact
that the larger the motor proteins, the larger is the area fraction $a_b$
introduced in \fig{fig:area} and therefore the more motors are on average
close enough to the microtubule to bind.

\begin{figure}[t!]
  \begin{center}
    \begin{tabular}{c@{\hspace{.04\linewidth}}c}
    \resizebox{.45\linewidth}{!}{\includegraphics{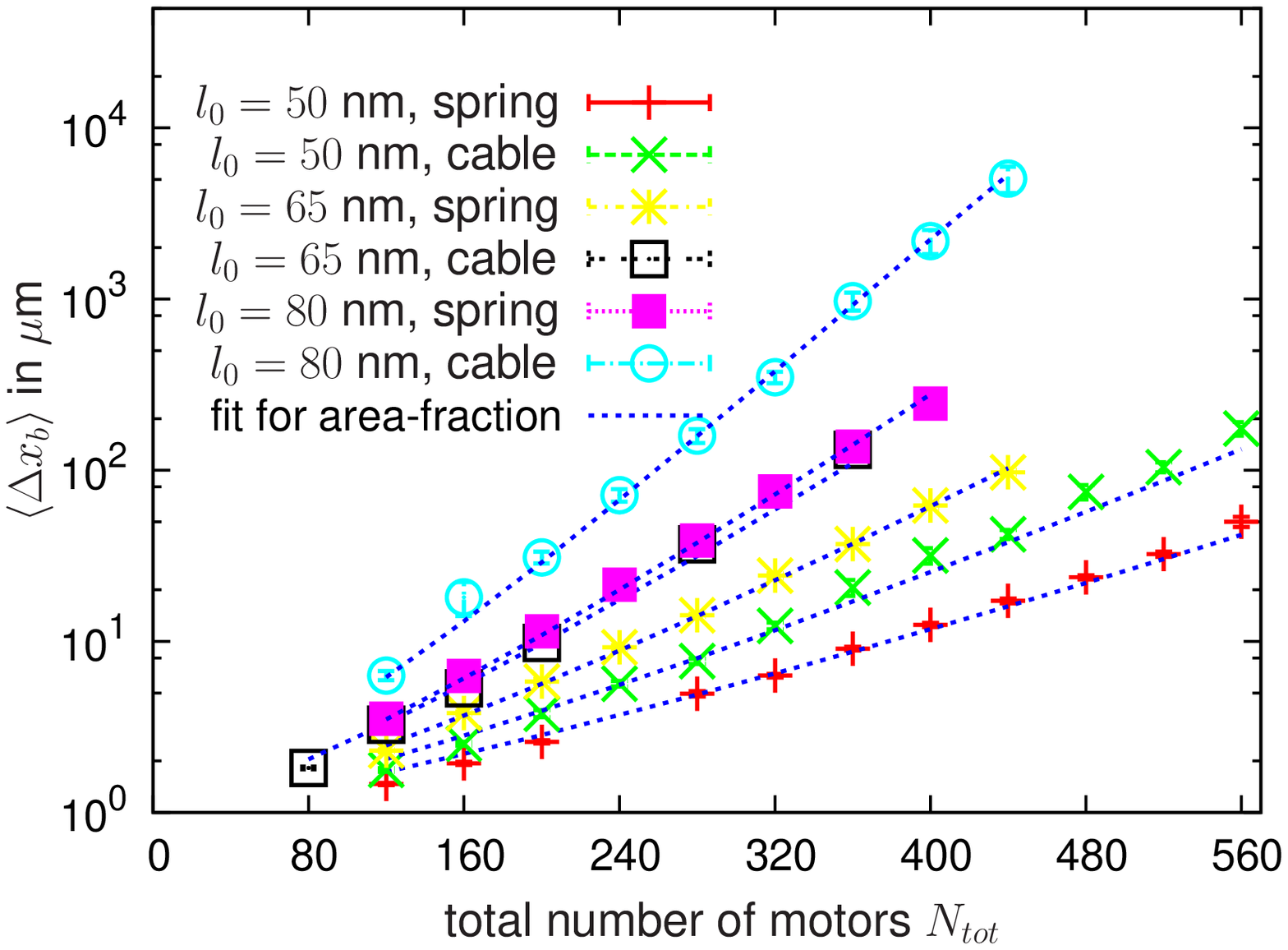}}&
    \resizebox{.45\linewidth}{!}{\includegraphics{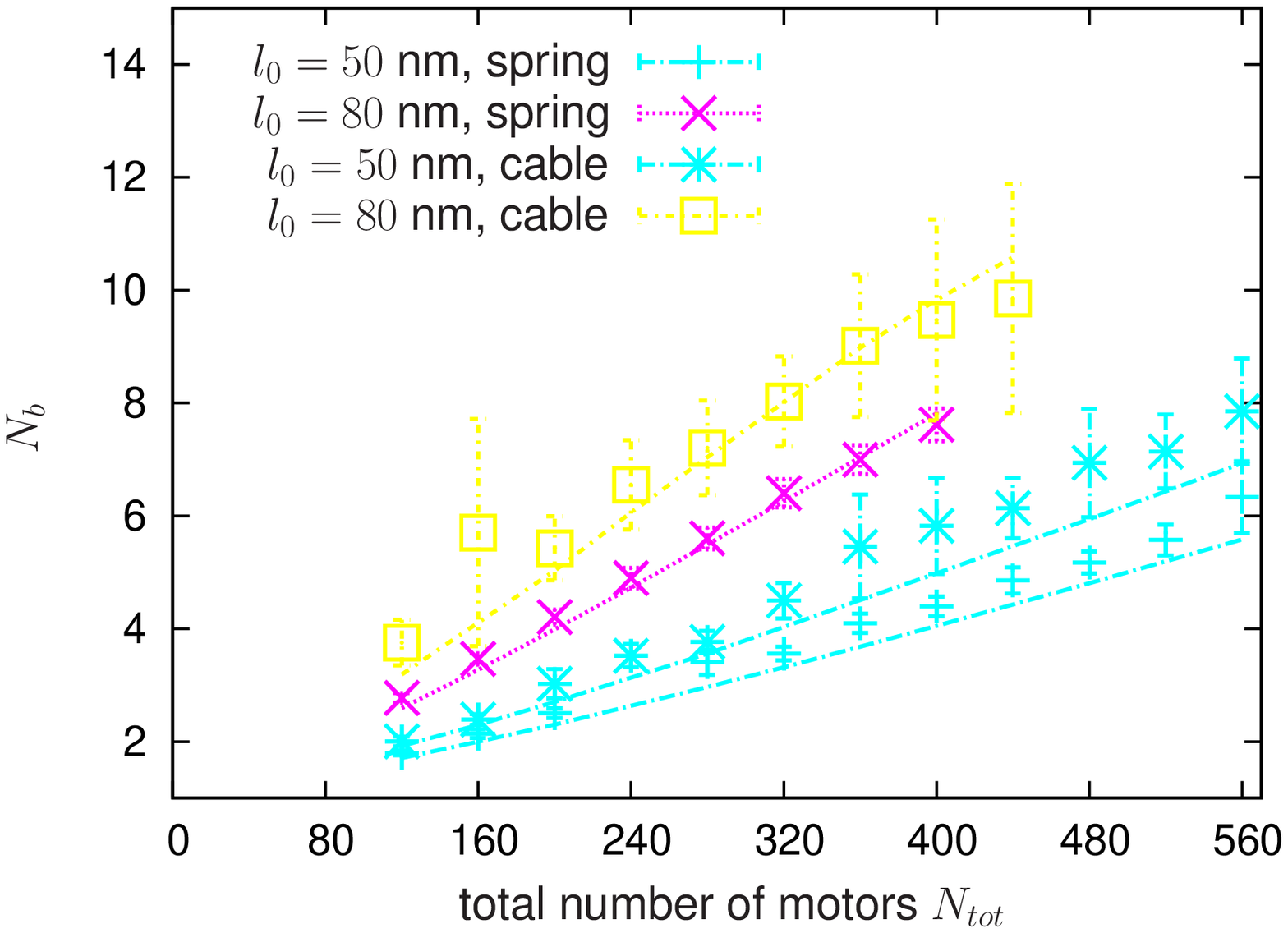}}\\ 
    (a) & (b) \\
    \end{tabular}
    \caption{(a) Mean run length $\dxb$ (data points with error
      bars) as a function of motors on the bead $N_{tot}$ obtained
      from adhesive motor dynamics. The lines are fits of
      \eq{meanmeanwalk} with respect to the area fraction $a_b$. (b) Mean
      number of bound motors $\mNb$ (data points with error bars). The lines
      are the values obtained from the Poisson-averaged mean number of
      bound motors $\mean{\mNb}_{Poisson}$ in \eq{meanmeanbound} using
      for $a_b$ the fit value from (a). (Parameters: $\pi_{ad} = 5,
      \epsilon_0 = 1$, $\lambda_s^0 = 125$, $\Delta t = 10^{-5}$, $\N
      \sim 10^4$.)
      \label{res:polymer:int1}}
  \end{center}
\end{figure}

\fig{res:polymer:int1}a shows the mean run length as a function of
$N_{tot}$ as obtained by numerical simulations of the transport process 
(points with error bars). For the motor stalk three different values of the
resting length $l_0 = 50,65,80$~nm are chosen and both the full-spring and the
cable model are applied for the force extension relation.
Similarly to what we have already seen for the run length distributions in
\fig{res:histos}, the larger the resting length $l_0$ the more motors can
simultaneously bind for given $N_{tot}$, and therefore the larger is the mean
run length. Furthermore, \fig{res:polymer:int1}a also demonstrates that
it makes a clear difference whether the motor stalk behaves like a full
harmonic spring or a cable. If the motor protein behaves like a cable
(semi-harmonic spring, \eq{eq:cable}) it exhibits force only if it is
stretched. The vertical component of this force always pulls the cargo towards
the MT. Thus, the mean height between the cargo and the MT (which determines
how many motors can bind at maximum) results from the interplay between this
force and thermal fluctuations of the bead.  In contrast, if the motor also
behaves like a harmonic spring when compressed, it once in a while may also
push the cargo away from the MT. This results in less motors being close
enough to the MT for binding than in the case of the cable-like behavior of
the motor stalk. Consequently, given the same $l_0$ and $N_{tot}$, the cargo
is on average transported longer distances when pulled by ``cable-like
motors''.

In order to apply the theoretical prediction for the mean run length of
a spherical cargo particle, \eq{meanmeanwalk}, we need to determine proper
values for the three parameters $a_b = n_b/N_{tot}, U, \Nbmax$. From the
simulations we measure the mean velocity of the sphere $U$. It turns out that
$U$ is up to 15~\% less than the maximum motor velocity $v_0$ due to
geometrical effects. Depending on the point where the motor is
attached on the sphere, some motor steps may result mainly in a slight rotation
of the sphere instead of translational motion of the sphere's center of mass
equal to the motor step length $\delta$. For $\Nbmax$ we choose the overall
measured maximum value from all simulation runs for given $N_{tot}$ and
polymer model of the motor. Then, we use the remaining parameter, the area
fraction $a_b$, as a fit parameter to the numerical results. The fits are done
using an implementation of the Marquardt-Levenberg algorithm from the
Numerical Recipes \cite{numerical:c}.  The resulting $a_b$ values are
summarized in \tab{tab:fitvalues}. The theoretical curves for those parameter
values are shown (dashed lines) in \fig{res:polymer:int1}a.
\begin{table}[t!]
\begin{center}
  \begin{tabular}{|c||c|ccc|}
    \hline $l_0$, motor-model & fit value for $a_b$& measured
    $\mean{h}$&$\rightarrow$& $a_b(\mean{h})$$\vphantom{\Big[}$\\
    \hline\hline $50$nm, spring \eq{eq:spring}&0.00211&7-14
    nm&$\rightarrow$& 0.0039-0.0034$\vphantom{\Big[}$\\ \hline $50$nm, cable
    \eq{eq:cable}&0.0026&4-11
    nm&$\rightarrow$&0.006-0.0055$\vphantom{\Big[}$\\ \hline $80$nm, spring
    \eq{eq:spring}&0.00403&8-14
    nm&$\rightarrow$&0.0082-0.0076$\vphantom{\Big[}$\\ \hline $80$nm, cable
    \eq{eq:cable}&0.00518&4-11
    nm&$\rightarrow$&0.0085-0.0079$\vphantom{\Big[}$\\ \hline
  \end{tabular}
  \caption{Obtained fit values for the area fraction $a_b$ for
    different $l_0$ and the two applied polymer models.  For
    comparison the area fraction which is obtained from the measured
    mean distance $\mean{h}$ is also displayed. $\mean{h}$ is measured
    for fixed $\N_{tot}$, the left boundary of the provided interval
    corresponds to the largest $N_{tot}$.\label{tab:fitvalues}}
\end{center}
\end{table}
The increase in $a_b$ for larger resting length and the cable model is
in excellent accordance with the above discussed expectation. An independent
estimate of the area fraction can be obtained by measuring the mean
distance $\mean{h}$ between cargo and MT and calculating the area
fraction $a_b$ as $a_b = a_b(\mean{h})$. For comparison those values
are also given in \tab{tab:fitvalues}. They turn out to be about 60 \%
larger than the values for $a_b$ obtained from the fit. This indicates
that the height of the sphere above the MT (and therefore also the
area fraction) is a fluctuating quantity that is not strongly peaked
around some mean value. Then, because of the non-linear dependence of
$a_b$ on $h$ we have in general $\mean{a_b(h)} \neq a_b(\mean{h})$.

\fig{res:polymer:int1}b shows the mean number of bound motors (the average is
obtained over all $\N$ simulation runs) as a function of $N_{tot}$
(symbols with error bars). The lines in \fig{res:polymer:int1}b are plots of
\eq{meanmeanbound} using the same parameters as for the correspondig lines in
\fig{res:polymer:int1}a. One recognizes that again the theoretical prediction
and the measured values match quite well. This means that on the level of
the mean run length and mean number of bound motors the Poission average that
was introduced in \sec{sec:poisson} works quite well, even though the number of
motors that are in range to the MT is not a constant during one simulation run
(cf. \sec{sec:inrange}).  The large error bars for the cable model data in
comparison to the spring model results partly from a poorer statistics (for
the $l_0 = 80$~nm cable simulations the number of runs is in the range of some
hundreds only). In addition, for cable-like motors the fluctuations in the
area fraction $a_b$ are much larger than for spring-like motors as repulsive
spring forces stabilize the distance between the sphere and the MT.  Therefore,
the width of the distribution density of the number of bound motors is larger
for the cable model than for the spring model. \fig{res:polymer:int1}b also
shows that for a bead radius of 1 $\mu$m, around 80 motors have to be attached
to the bead, otherwise binding and motor stepping become unstable because
there are less than two motors left in the binding range.

\begin{figure}[t!]
  \begin{center}
    \resizebox{.5\linewidth}{!}{\includegraphics{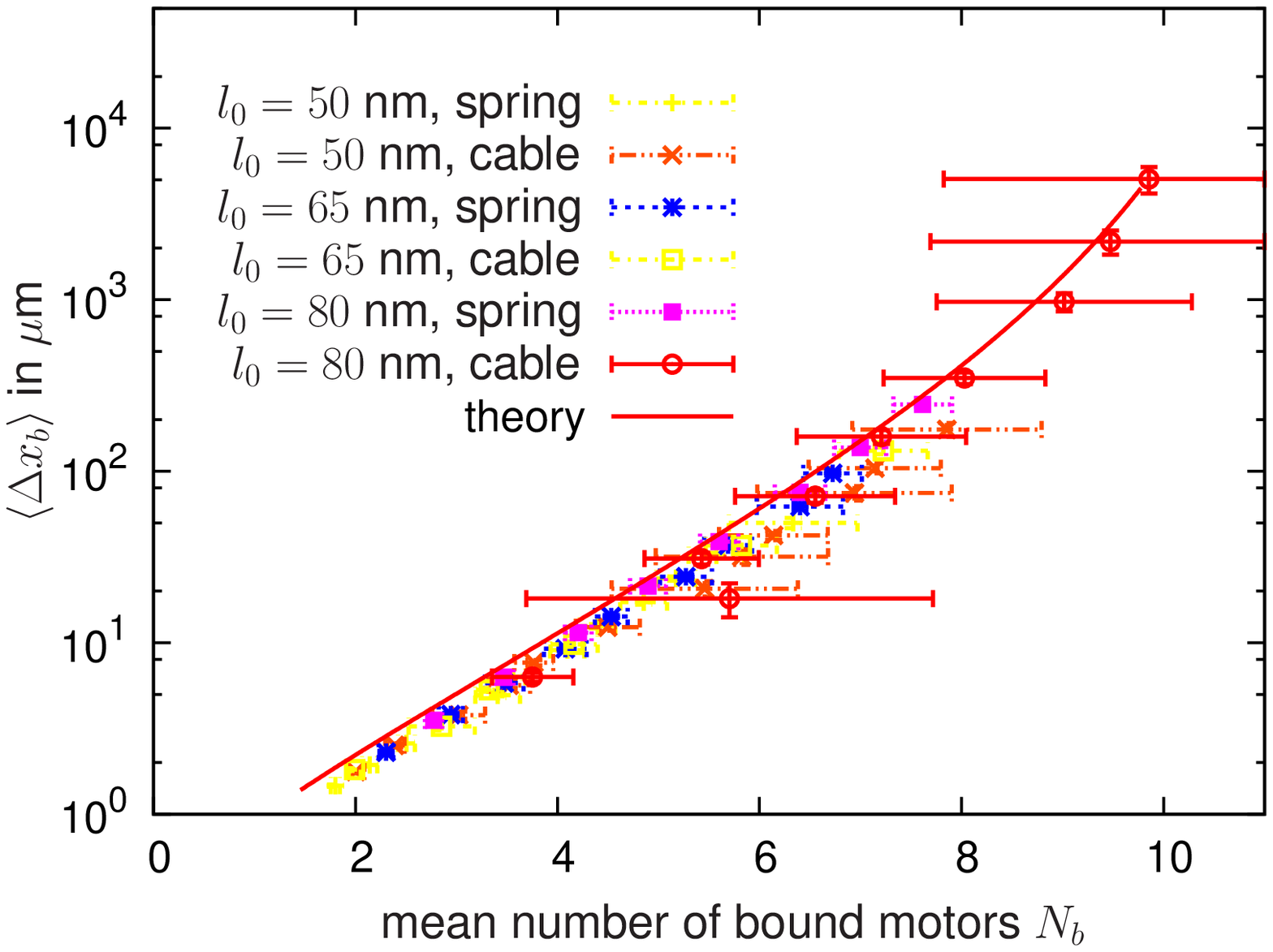}}
      \caption{Combination of data from \fig{res:polymer:int1}a,b for the
	resting lengths $l_0 = 50,65,80$~nm. $\dxb$ is shown as a function of
	$\mNb$. For the dashed line (theory) \eq{meanmeanwalk} and
	\eq{meanmeanbound} were combined with a truncation of the sums at
	$N_{m,max} = 18$. (Parameters: $\pi_{ad} = 5$~s, $\epsilon_0 = 1$~s,
	$\lambda_s^0 = 125$~s, $\Delta t = 10^{-5}/\epsilon_0$, $\N \sim
	10^4$.)
      \label{res:walk_bound}}
  \end{center}
\end{figure}
In \fig{res:walk_bound} the simulation results of \fig{res:polymer:int1}a,b
are combined into one plot. Here we show the measured mean run length $\dxb$ as a
function of the mean number of bound motors $\mNb$.  All curves
collapse on one master curve that can be parametrized by $n_b = a_b N_{tot}$,
i.\,e., the product of the fit parameter $a_b$ and the total number of motors
on the sphere. The fact that all data points turn out to lie on one master
curve again demonstrates the good applicability of the theoretical predictions
to the simulation results. The curve shown in \fig{res:walk_bound} has a
positive curvature in the semi-logarithmic plot. This turns out to be an
effect of the finite truncation of the sums in \eq{meanmeanwalk} and
\eq{meanmeanbound}.
 
\subsection{Distribution of motors in binding range}
\label{sec:inrange}

\begin{figure}[t!]
  \begin{center}
    \begin{tabular}{c@{\hspace{.04\linewidth}}c}
    \resizebox{.45\linewidth}{!}{\includegraphics{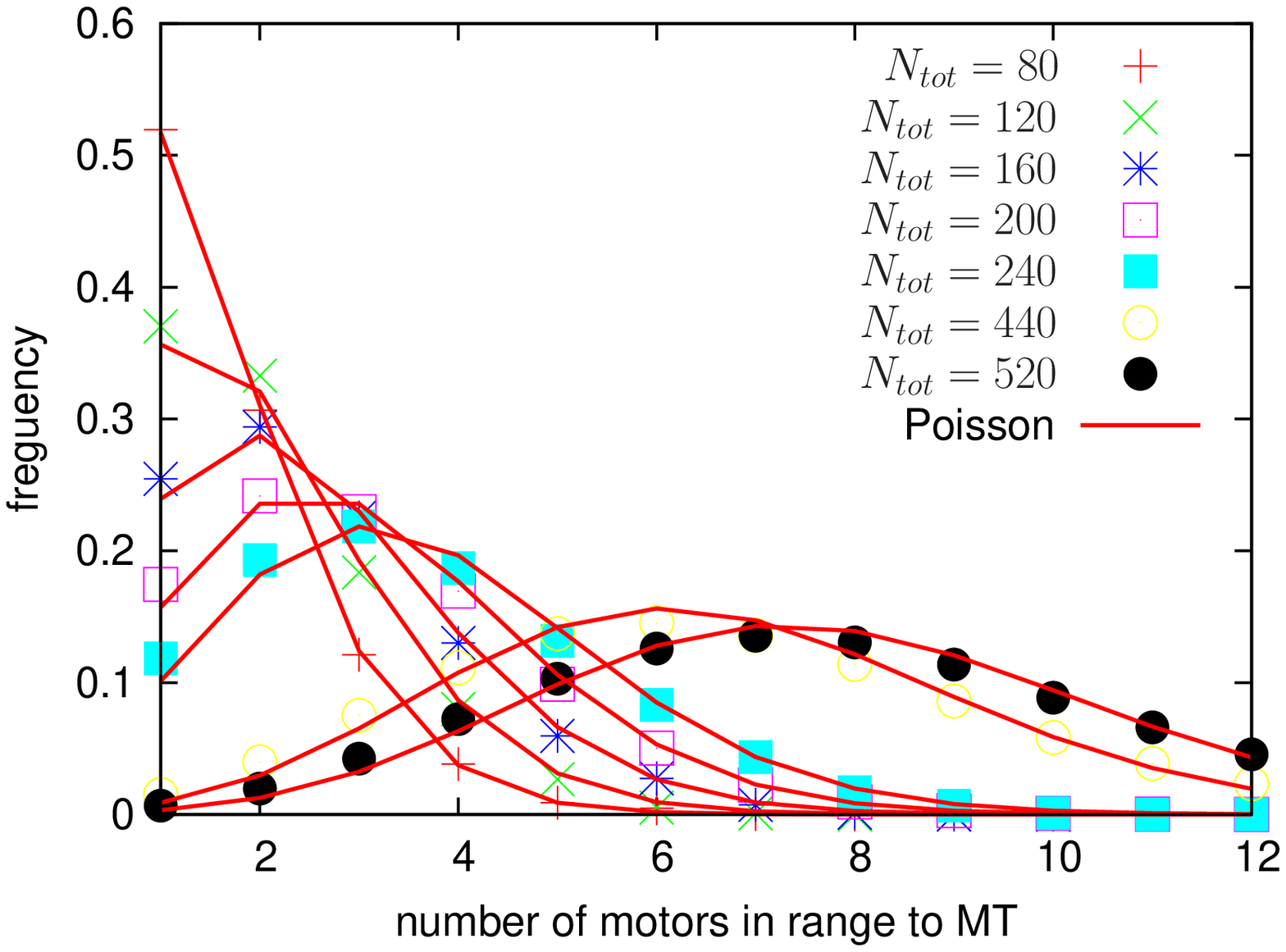}}&
    \resizebox{.45\linewidth}{!}{\includegraphics{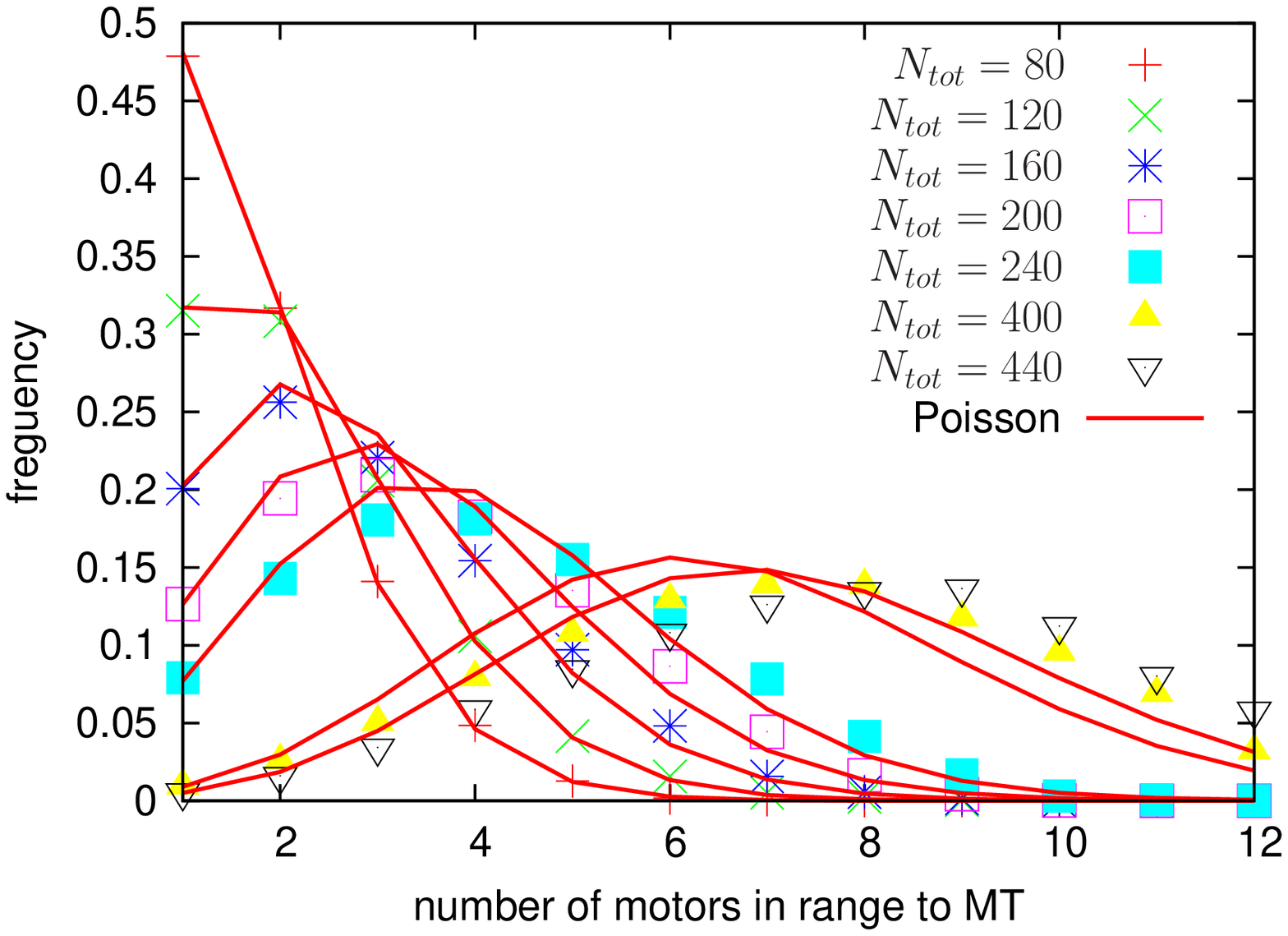}}\\ 
    (a) & (b) \\
    \end{tabular}
    \caption{Histograms for the number of motors that are in binding range to
      the MT. Symbols refer to simulation results for different values of the
      total number of motors on the sphere $N_{tot}$. Lines are Poisson
      distributions with mean value that is propotional to $N_{tot}$.
      (a) Resting length $l_0 = 50$~nm, spring model, \eq{eq:spring}. 
      (b) Resting length $l_0 = 50$~nm, cable model, \eq{eq:cable}. 
      (Parameters: $\pi_{ad} = 5,
      \epsilon_0 = 1$, $\lambda_s^0 = 125$, $\Delta t = 10^{-5}/\epsilon_0$, $\N
      = 2\cdot 10^4$.)
      \label{fig:inrange}}
  \end{center}
\end{figure}

For the evaluation of the numerical data in \sec{sec:walking} we assumed that
the number of motors that are in binding range to the MT are constant in
time and that this number is drawn from a Poisson distribution for every
individual run. We now further examine this aspect in order to see to what
extend this assumption is fulfilled. First, we measure directly the distribution
of the number of motors $n_{MT}$ in binding range to the MT. For this we count
$n_{MT}$ at every numerical time step during one simulation run and repeat
this for a large ensemble of runs ($\N \sim 10^4$). Thus, the histograms
(i.\,e., approximately the probability distributions we are looking for) that
are obtained in this way for the relative frequency of $n_{MT}$ are not based
on the assumption of constant $n_{MT}$ for every single run.

\fig{fig:inrange} shows examples of such histograms (symbols) for a
series of different values of the total number of motor coverage
$N_{tot}$.  For \fig{fig:inrange}a we used the spring model for the
motor polymer, \eq{eq:spring}, and for \fig{fig:inrange}b the cable
model, \eq{eq:cable}. In both cases the resting length of the motor
protein is $l_0 = 50$~nm.  In addition, \fig{fig:inrange} also
displays probability distributions (solid lines) that are obtained
from the Poission distribution given that at least one motor is in
binding range, i.\,e., $p(n) = \mu^n e^{-\mu}/n!(1-e^{-n})$, with mean
value $\mu$ given by $\mu = \mu_0N_{tot}$ that can be parametrized by
some variable $\mu_0$.  The parameter $\mu_0$ was chosen to be 0.015
and 0.0165 for the spring and cable model, respectively, by matching
the Poisson distribution to the simulation data.  For the spring model
(\fig{fig:inrange}a), the fit is excellent for all values of
$N_{tot}$.  One must note however that these distributions are not
given by \eq{poisson} as indicated by the fact that the parameter
$\mu_0$ is much larger than the area fraction $a_b$ determined in the
previous section. Instead one needs to account for the correct
weighting factors from the different run lengths.  If for the moment
we consider the number $n_{MT}$ to be a constant during one run, then
the distribution \eq{eq:probn} can be considered to be a useful
estimate for the data in \fig{fig:inrange}. Taking the limit $\Nbmax
\rightarrow \infty$ in \eq{eq:probn} we obtain
\begin{align}
  \label{eq:inrange}
  \frac{(a_b(1+\pi_{ad}/\epsilon_0))^{n_{MT}} - a_b^{n_{MT}}}{e^{a_b\pi_{ad}/\epsilon_0} -
  1}\frac{e^{-a_b}}{{n_{MT}}!} \approx \frac{b^{n_{MT}}}{{n_{MT}}!}e^{-b}, 
\end{align}
with $b = a_b(1+\pi_{ad}/\epsilon_0)$. Thus, the result is
approximately---except for very small $n_{MT} = 1,2$---again a Poisson
distribution with parameter $b$. With $\pi_{ad}/\epsilon_0 = 5$ and $a_b =
0.0021$ from \tab{tab:fitvalues} we get $b = 0.013$ which is very close
to the value $\mu_0 = 0.015$ used in \fig{fig:inrange}a.

For the cable model (\fig{fig:inrange}b), the Poisson fit using a single value for $\mu_0$ works well only for the smaller values
of $N_{tot}$. For large $N_{tot}$ the data points cannot be fitted by a
Poisson distribution. It rather turns out that the ratio of mean
value and standard deviation becomes less than one in these cases.

\begin{figure}[t!]
  \begin{center}
    \resizebox{.45\linewidth}{!}{\includegraphics{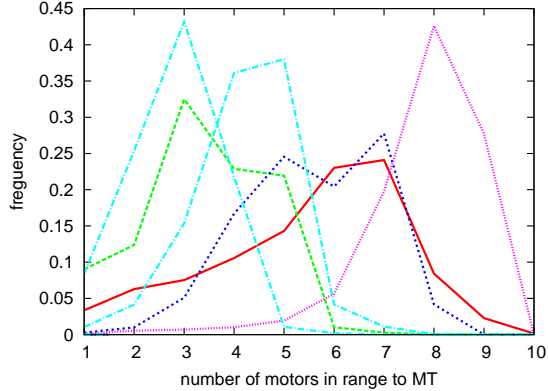}}
      \caption{The relative frequencies of the number of motors that are in
	binding range to the MT during a single run are shown for six
	different sample runs. For the motor protein the cable model with
	resting length $l_0 = 50$~nm is used. (Parameters: $N_{tot} = 400$,
	$\pi_{ad} = 5$~s, $\epsilon_0 = 1$~s, $\lambda_s^0 = 125$~s, $\Delta t
	= 10^{-5}/\epsilon_0$).
      \label{fig:singleinrange}}
  \end{center}
\end{figure}
Finally, we shall note that despite the good agreement between the
simulation data and the the estimate from \eq{eq:inrange}, which was
based on the assumption that $n_{MT}$ is constant during one run,
$n_{MT}$ is in fact not constant, but a fluctuating quantity. The
fluctuations result partly from thermal fluctuations of the height and
orientation of the sphere and partly from orientation changes of the
sphere that are induced by motor forces. \fig{fig:singleinrange} shows
a few sample histograms for the frequency that $n_{MT}$ motors are in
binding range to the MT, i.e.\ either bound to the MT or unbound
within the binding range, during a single run. These examples clearly
show that $n_{MT}$ takes different values during one run.

\subsection{Escape rate distributions}
\label{sec:escaperate}

\begin{figure}[t!]
  \begin{center}
    \resizebox{.45\linewidth}{!}{\includegraphics{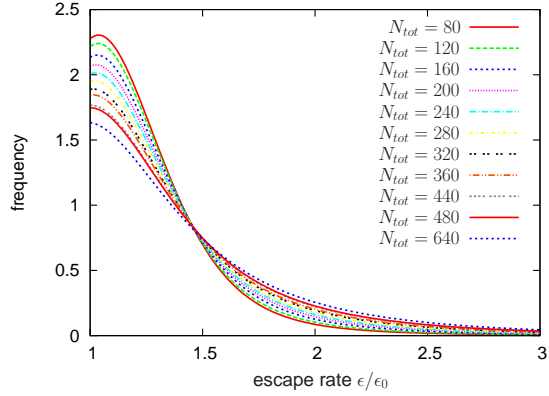}}
    \caption{Measured probability distribution density for the escape rate
      $\epsilon$ given $\epsilon > \epsilon_0$. The data was obtained for
      different values of $N_{tot}$. For the motor proteins the full spring
      model, \eq{eq:spring}, with resing length $l_0 = 50$~nm was used.
      (Parameters: $\pi_{ad} = 5,
      \epsilon_0 = 1$, $\lambda_s^0 = 125$, $\Delta t = 10^{-5}/\epsilon_0$, $\N
      = 2\cdot 10^4$.)
      \label{fig:escapehisto}}
  \end{center}
\end{figure}

Diffusive motion of the cargo sphere directly influences the length of the
pulling motor proteins and therefore also the escape rate $\epsilon$. The
dependence of the escape rate on the motor length $x$ for $x > l_0$ is
obtained by inserting the polymer model \eq{eq:cable} and \eq{eq:spring},
respectively, into the Bell equation, \eq{eq:bell}. For $x \leq l_0$ the
escape rate is given by $\epsilon = \epsilon_0$. For low viscous friction we
assume $x$ to be distributed according to a Gaussian distribution. 
Then, we expect the probability distribution density $\tilde
p(\epsilon)$ for the escape rate $\epsilon$ to be given by a log-normal
distribution density
\begin{align}
  \label{eq:lognorm}
  \tilde p(\epsilon) = \frac{1}{
    b\epsilon}\exp{\left(\frac{F_d^2(\ln(\epsilon) - \ln(\bar \epsilon))^2}{2
	k_B T \kappa^*}\right)}.
\end{align}
Here, $\bar\epsilon$ denotes the escape rate associated with the mean motor
length in the extended state,
$\kappa^*$ is an effective spring constant that depends, e.\,g. on the number
of pulling motors, and $b$ is a normalization constant that is obtained from
the condition
\begin{align*}
  \int_{\epsilon_0}^\infty \tilde p(\epsilon) d\epsilon \stackrel{!}{=}1.
\end{align*}
\fig{fig:escapehisto} shows the measured probability density for 
$\epsilon > \epsilon_0$ and different values of $N_{tot}$. It turns out that the
log-normal distribution in \eq{eq:lognorm} matches well to the
measured data (not shown). Fitting \eq{eq:lognorm} to the data for the effective spring
constant $\kappa^*$ it turns out that $\kappa^*$ increases with increasing 
$N_{tot}$. This makes sense and illustrates that for several motors pulling
the motors behave as a parallel cluster of springs.

The agreement of \eq{eq:lognorm} with the simulation data for the
escape rate distributions suggests that the main source of force
acting on the motors and increasing the unbinding rate is due to
thermal fluctuations of the micron-sized bead.  This is different from
what has been reported in experiments with nano-scaled two-motor
complexes \cite{rogers09}.  There it has been argued that the forces
between the two motors arising from their stochastic stepping lead to
an increased unbinding rate of these motors.

\subsection{Cargo transport against high viscous friction}
\label{sec:velocity}

Except for the single motor simulations of \sec{sec:single}, all
simulation data discussed so far were obtained for a viscosity of
$1$~mPa~s, corresponding to a water-like solution. We mentioned in
\sec{sec:mdynamics} that load sharing between several motors may lead
to cooperative effects at high viscous friction. We now analyze this
further by performing simulations at viscosities much larger
than that of water (i.\,e., when the viscous friction on the bead is
comparable to the internal friction of the motor protein). To do this
we need to measure the velocity of the bead depending on the number of
pulling motors.  Because of the nature of the stochastic process
describing the position of the cargo the instantaneous cargo velocity
is however not well defined \cite{gardiner:85}. Therefore, in order to
measure the cargo velocity $U$ we have to average over some time
interval $\Delta \bar t$.  If no motors were pulling the velocity
distribution density is given by a Gaussian,
\begin{align}
  \label{eq:veldistr}
  p(U,\Delta \bar t) = \sqrt{\frac{\Delta \bar t}{4\pi D}}
  e^{-U^2 \Delta \bar t/4 D},
\end{align}  
with diffusion constant $D = k_B T_a \mu_{xx}^{tt}$
(cf. \sec{sec:stokesian}). Furthermore we assume a constant height of the
sphere so that the mobility coefficient $\mu_{xx}^{tt}$ is a constant in time.
The width of the distribution density, \eq{eq:veldistr}, is the smaller the
larger $\Delta \bar t$ is. So in order to suppress fluctuation effects it
seems appropriate to average over a large time interval $\Delta \bar t$. On the
other hand the number of pulling motors changes with time because of binding
and unbinding. Thus, in order to measure the velocity given a certain number
of motors $\Delta \bar t$ should not be too large in order to get enough such
events. Here we choose $\Delta \bar t = 0.02$~s which corresponds to a typical
camera resolution of 50~Hz.

\begin{figure}[t!]
  \begin{center}
    \begin{tabular}{c@{\hspace{.04\linewidth}}c}
      \resizebox{.46\linewidth}{!}{\includegraphics{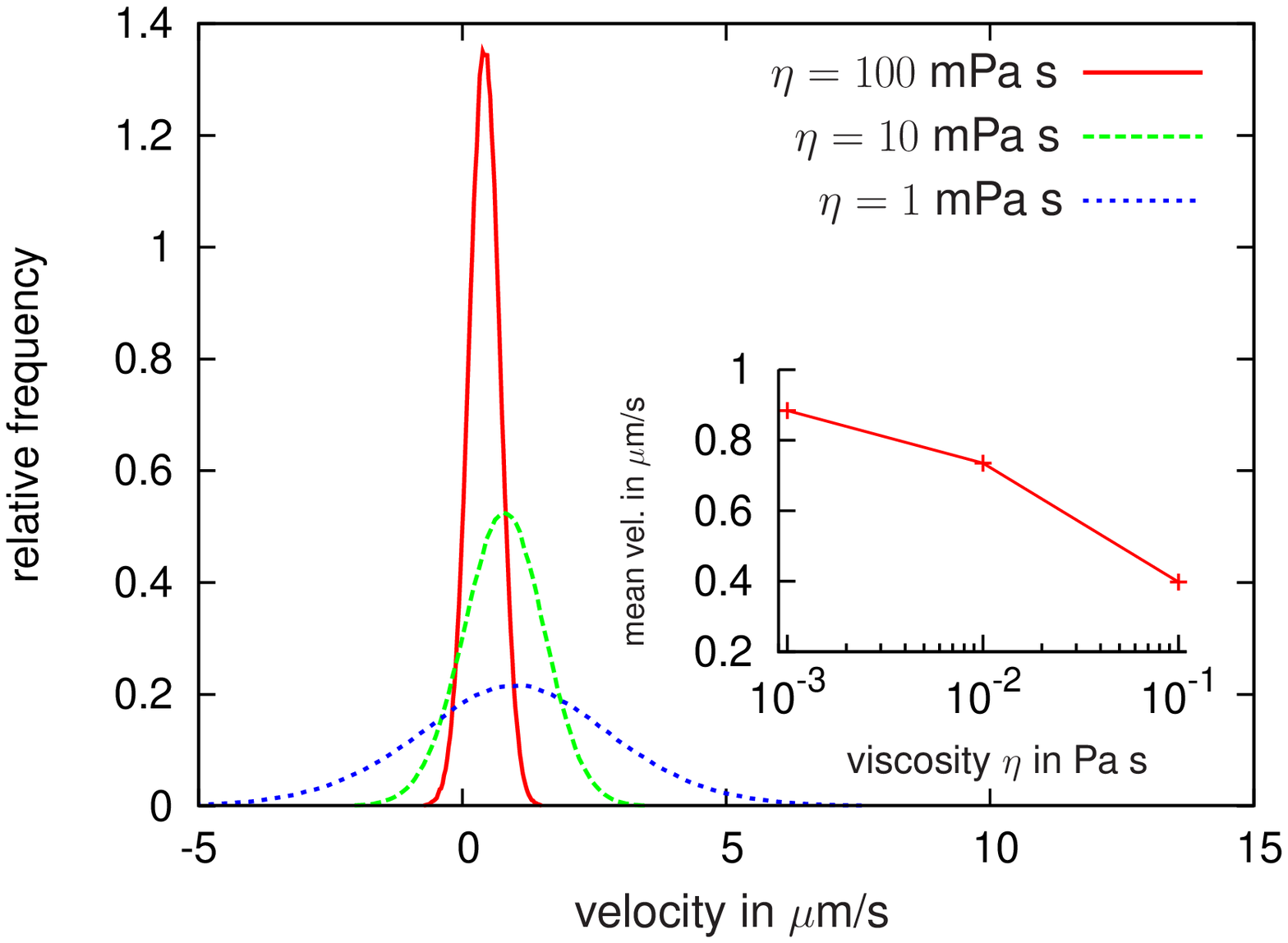}}&
      \resizebox{.46\linewidth}{!}{\includegraphics{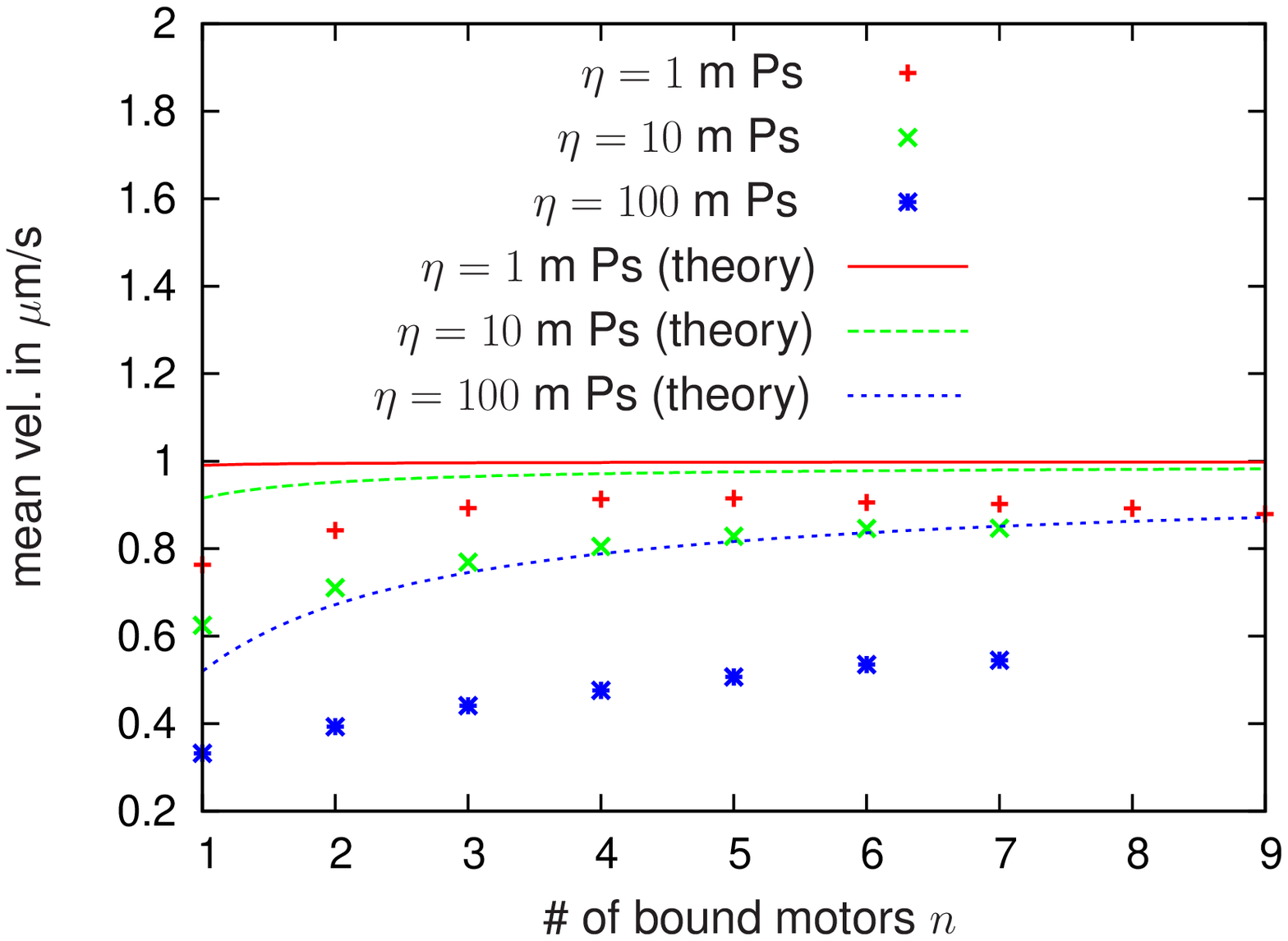}}\\
      (a) & (b)\\
      \resizebox{.46\linewidth}{!}{\includegraphics{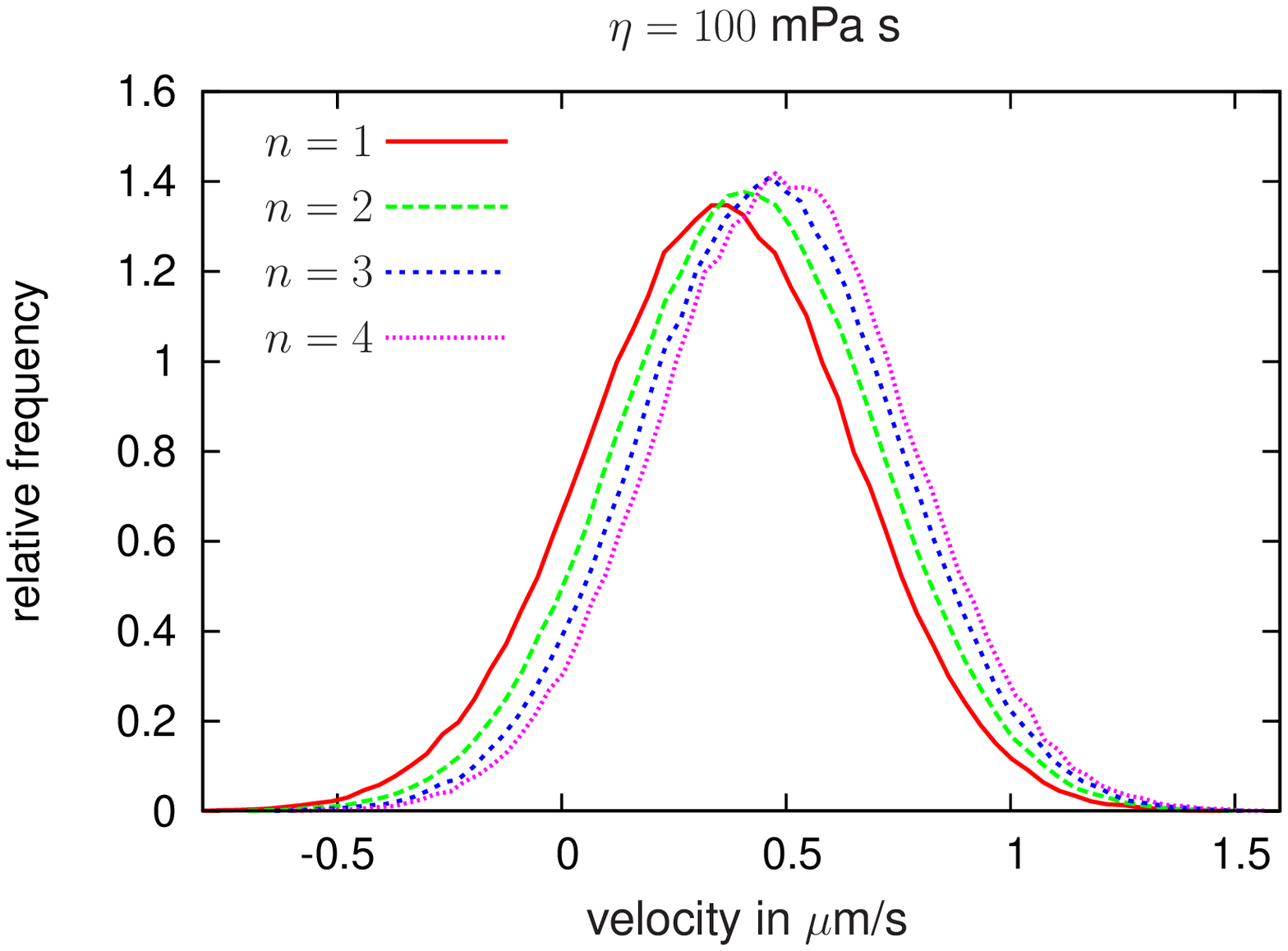}}&
      \\
      (c)\\
    \end{tabular}
    \caption{(a) Propability density of the cargo particle's velocity $U$ that
      is obtained by averaging over a time interval of $\Delta \bar t =
      0.02$~s for different values of the viscosity $\eta$. The inset shows
      the mean velocity as a function of the $\eta$.  (b) The mean velocity of
      the bead given that $n$ motors are simultaneously pulling is plotted as
      a function of $n$ (symbols). For comparison the theoretical expectation
      according to \eq{eq:velratio} is plotted, too (lines). (c) For high viscosity $\eta = 100$~mPas
      the conditional velocity distribution density given that a certain number of motors is pulling is plotted.
       (The full
      harmonic spring model was used; parameters: $l_0 = 80$~nm, $N_{tot} =
      200$, numerical time step $\Delta t = 10^{-5}$, other parameters as in
      \tab{tab:parameter}.)}\label{fig:meanvel}
\end{center}
\end{figure}
\fig{fig:meanvel}a shows the measured velocity distributions for three
different values of the viscosity, $\eta = 1,10,100$~mPa~s. In the
inset of \fig{fig:meanvel}a the mean velocity is plotted as a function
of the viscosity. It turns out that shifting the distribution
\eq{eq:veldistr} by the corresponding mean velocity, the single peaked
function $p(U,\Delta \bar t)$ fits qualitatively well to the
distribution shown in \fig{fig:meanvel}a, especially the dependence of
the width of the distributions on $\eta$ is correctly predicted by
\eq{eq:veldistr}. Also a decrease of the mean velocity of the cargo
particle is observed with increasing viscosity resulting from the
increased frictional load. It must be emphasized that only a single
peak is observed in \fig{fig:meanvel}a, even though the bead
velocity is expected to depend on the number of pulling motors, which
should lead to multiple distinct peaks \cite{klumpp:05}. One reason
that we do not observe multiple peaks is the small value of the time
interval $\Delta\bar t$, which leads to broad peaks with a peak width
governed by diffusion of the bead (cf.\ \eq{eq:veldistr}) and thus
makes it impossible to separate different peaks. Using larger values
of $\Delta \bar t$ leads to smaller peak widths, but also to poorer
statistics as less measurement points are obtained, so that again
distinct peaks cannot be resolved. Even if we do not use a constant
time interval, but average over the variable time intervals between
two subsequent changes in the number of bound motors \cite{hill:04},
distinct peaks are very hard to separate (not shown).  This does
however not mean that the bead velocity is independent of the number
of pulling motors. Indeed if we plot the conditional velocity
distribution calculated over all intervals in which the bead is pulled
by a certain fixed number of motors, we see a clear shift in the
average velocity (\fig{fig:meanvel}c). This shift is however masked by
the width of the distributions in \fig{fig:meanvel}a.

In \fig{fig:meanvel}b the average of all measured velocities given
that exactly $n$ motors are pulling is plotted as a function of $n$,
again for the three different viscosities $\eta = 1,10,100$~mPa~s.
For $\eta = 1$~mPa~s the viscous friction for the bead is about
$1/\mu_{xx}^{tt} \approx 5\cdot 10^{-8}$~Ns/m. The internal friction
of the motor is $1/\mu_m = F_s/v_0 \approx 6\cdot 10^{-6}$~Ns/m,
i.\,e., about two orders of magnitude larger than $1/\mu_{xx}^{tt}$.
According to the analysis at the end of \sec{sec:mdynamics} we
therefore expect that the mean velocity is independent of $n$ if all
motors equally share the load. The numerical data in
\fig{fig:meanvel}b shows that the mean velocity exhibits a weak
dependence on $n$ with a maximum at about $n = 4,5$.  At the higher
viscosities the numerical results show that the mean bead velocity
increases with increasing $n$, which indicates that the motors share
the load. The simulation data however deviate clearly from the
estimate given by \eq{eq:velratio}, which is indicated by the lines in
\fig{fig:meanvel}b. This discrepancy indicates that the load is not
shared equally among the motors or that only a subset of the bound
motors are actually pulling the bead.  Besides geometrical effects one
reason why this is the case is that the escape rate $\epsilon_0$ is
rather high and at high frictional load is even further increased
making the lifetimes of motors in the pull state rather short. Then,
if a new motor binds to the MT often another motor detaches already
before a stationary state is reached in which all the motors equally
share the load.

\begin{figure}[t!]
  \begin{center}
    \begin{tabular}{c@{\hspace{.04\linewidth}}c}
      \resizebox{.46\linewidth}{!}{\includegraphics{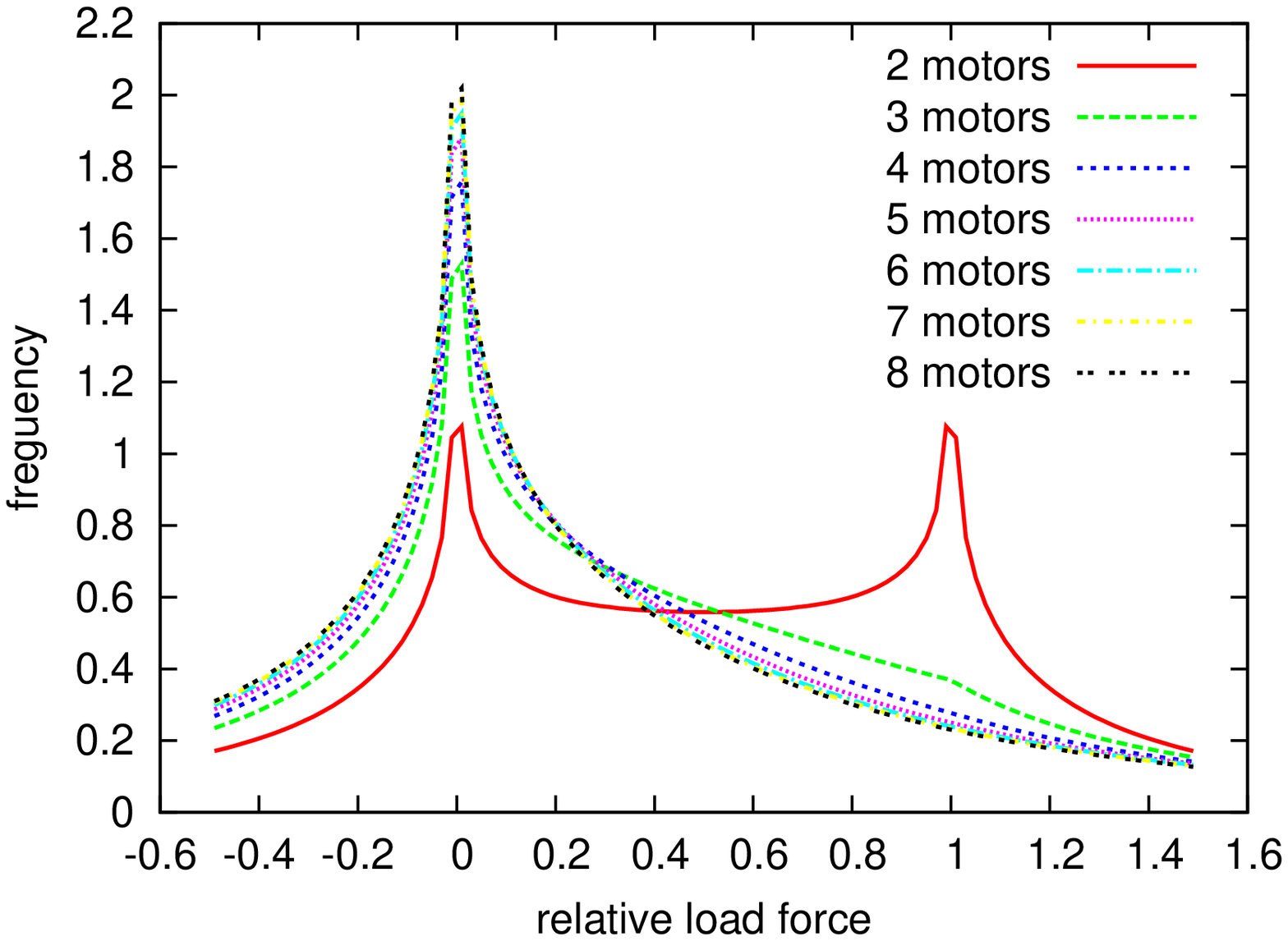}}&
      \resizebox{.46\linewidth}{!}{\includegraphics{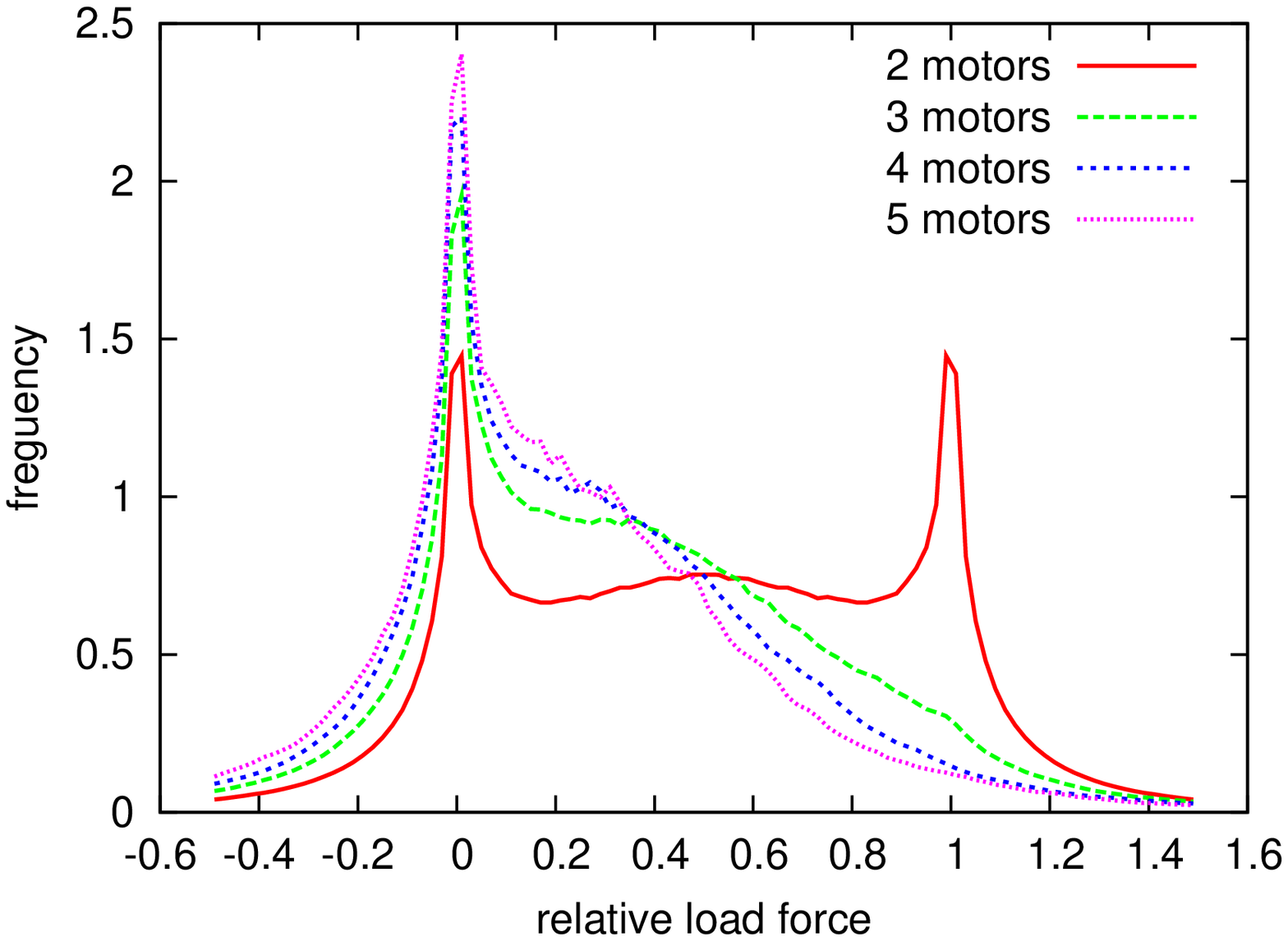}}\\
      (a) & (b) \\
      \resizebox{.46\linewidth}{!}{\includegraphics{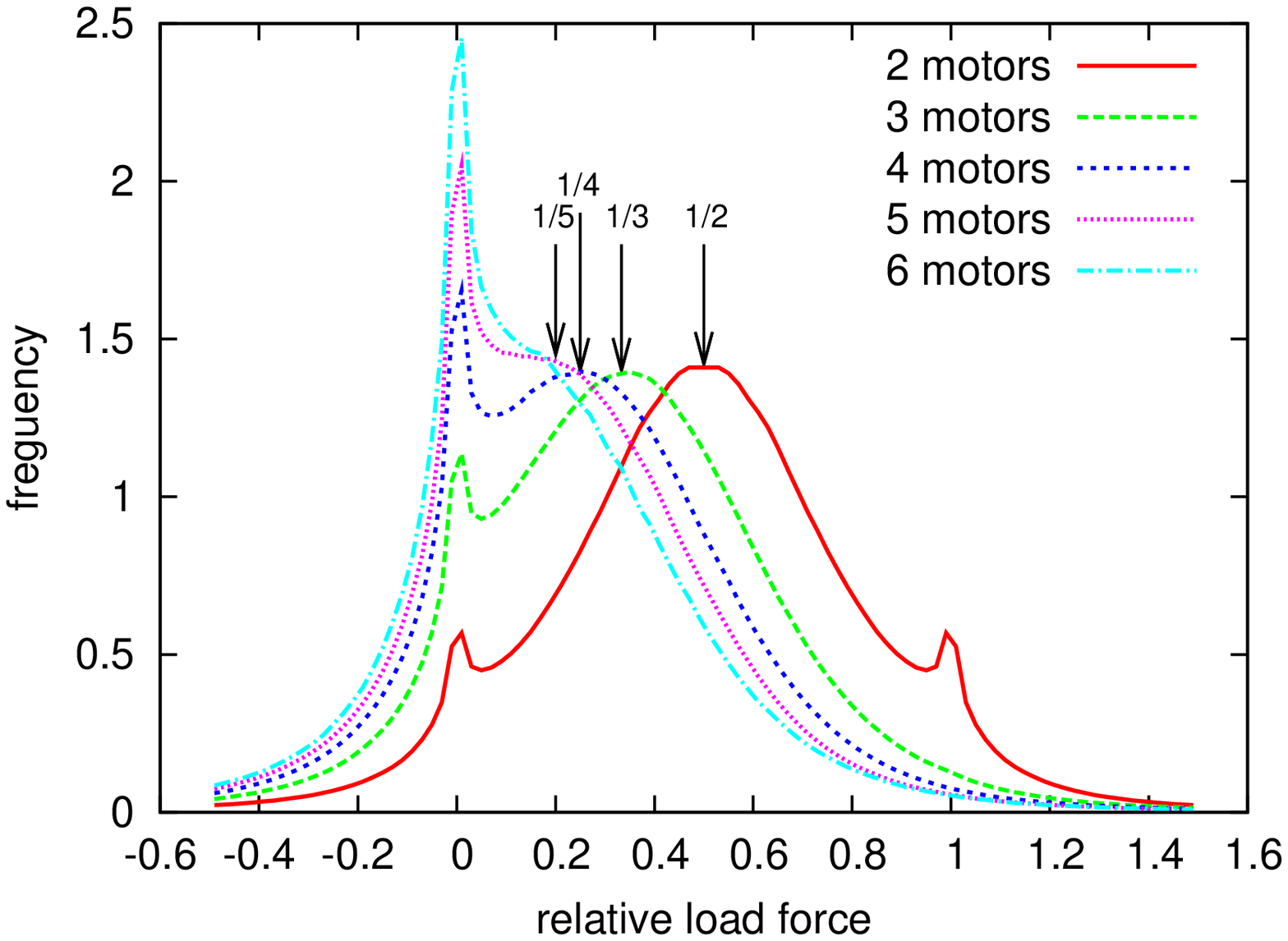}}&
      \resizebox{.46\linewidth}{!}{\includegraphics{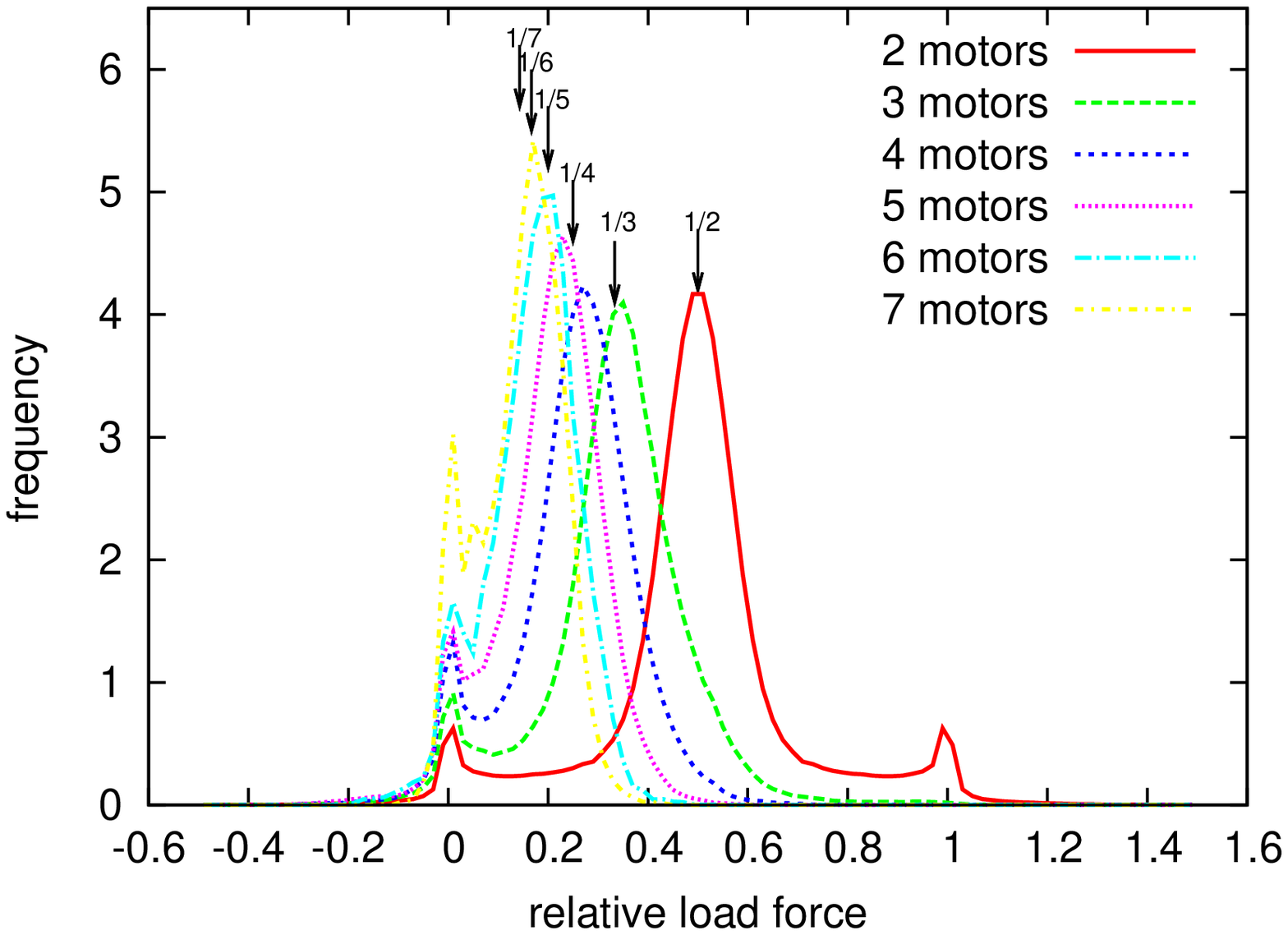}}\\
      (c) & (d)\\
    \end{tabular}
    \caption{Frequencies of the motor forces in walking direction relative to
      the total load force on the cargo particle for different numbers of
      pulling motors.
      (a) Viscosity $\eta = 1$~mPa~s, escape rate $\epsilon_0 = 1$~s$^{-1}$. 
      (b) $\eta = 100$~mPa~s, $\epsilon_0 = 1$~s$^{-1}$. 
      (c) $\eta = 100$~mPa~s, $\epsilon_0 = 0.1$~s$^{-1}$.
      (d) $\eta = 1000$~mPa~s, $\epsilon_0 = 0.1$~s$^{-1}$.
      For the other parameters the same values as in 
      \fig{fig:meanvel} were used.}\label{fig:forcehistos}
\end{center}
\end{figure}
In order to investigate the last point in more detail we consider
explicitely how the force is typically distributed among the pulling
motors. For this we count the number $n$ of motors attached at each
numerical time step and measure the force experienced by each motor in
the direction along the microtubule. For a given number $n$, $n$ such
motor forces can be measured, $F_{m,x}^{(i)}, i = 1,\ldots,n$. To
suppress effects from fluctuations in the overall load we then
calculate the reduced forces $f_n := F_{m,x}^{(i)}/\sum_{i=1}^n
F_{m,x}^{(i)}$. Given the histograms for these quantities measured
over many time steps and simulation runs we obtain an approximation
for the probability distribution density of the relative load of the
motors.  \fig{fig:forcehistos} shows results for such histograms
obtained at different viscosities and two different values of the
unstressed escape rate $\epsilon_0$. \fig{fig:forcehistos}a shows the
results for $\eta = 1$~mPa~s and $\epsilon_0 = 1$~s$^{-1}$. One can
see some symmetries that result from the definition of the reduced
forces. This is especially emphasized for the case of $n = 2$ pulling
motors. The distribution density of the reduced forces has a mirror
symmetry about the value $f_2 = 0.5$. Besides this artefact it turns
out that for $n > 2$ the distribution densities are strongly peaked at
zero force. Thus, for low viscous friction the overall load results
mainly from fluctuations. Such loads are typically experienced by a
single motor only, whereas the remaining motors are more or less
un-stretched.  In \fig{fig:forcehistos}b the viscosity is 100 times
larger than that of water which causes an appreciable load on the
pulling motors. However, load sharing effects are hardly visible
except for the maxima around $f_2 = 1/2$ and $f_3 = 1/3$, which are
rather broad and thus hard to resolve. On the other hand there is
still a strongly peaked maximum at $f_n = 0$. This somewhat surprising
observation is due to the rather high escape rate, which is even
increased by the load force (cf. \eq{eq:bell}). Therefore, the binding
time of the motors is shortend. On the other hand motors that bind to
the MT are initially unstressed (i.\,e., carry zero load) in our
model. Thus they always contribute to the $f_n = 0$ peak and may
already escape from the MT before the load is shared equally by all
pulling motors.  When the escape rate is reduced to $\epsilon_0 =
0.1$~s$^{-1}$ as done for \fig{fig:forcehistos}c,d clearly visible
maxima around $f_n = 1/n$ appear in the histograms that indicate that
the load is equally shared by the active motors. In
\fig{fig:forcehistos}d where we used the extremely high value of
$10^{3}$~mPa~s for the viscosity, these peaks are very pronounced. The
arrows in \fig{fig:forcehistos}c,d indicate the relative force values
$1/i, i = 2,3,\ldots$. It turns out that the peaks are not exactly
located at these values which is again due to the binding and
unbinding process of the motors.

In summary, we found that at high loads the pulling motors tend to
arrange in such a way that the total load is equally shared amongst
them. However, for typical escape rate values of kinesin-like motors
this process often takes more time than the lifetime of a state of a
certain number of motors bound to the MT lasts, thus preventing
cooperativity in the sense of equal load sharing.

\section{Discussion and Outlook}
\label{sec:discussion}

The main purpose of this paper is to introduce a novel algorithm
called \textit{adhesive motor dynamics} as a means to study the
details of motor-mediated cargo transport.  Our algorithm is an
extension of existing adhesive dynamics algorithms developed to
understand the physics of rolling adhesion \cite{hammer:92,korn:07a}.
Basically our method allows to simulate the motion of a sphere above a
wall including hydrodynamic interactions and diffusive motion by
numerically integrating the Langevin equation, \eq{langevin-euler}. In
addition, motor-specific reactions such as binding to the microtubule
and stepping are modeled as Poisson processes and then
translated into algorithmic rules. The parameters and properties by
which the motors are modeled are based on results of single-molecule
experiments with conventional kinesin. A first favorable test for the
algorithm was provided by measuring the force velocity--relation at
different viscosities and external load forces and by comparing the
results to the input data as done in \sec{sec:single}.

Next we measured the run length and the mean number of bound motors as
a function of the total number $N_{tot}$ of motors attached to the
sphere. The same quantities have been previously calculated based on a
one-step master equation model \cite{klumpp:05}. However, this has
been done as a function of a fixed number of motors that are in
binding range to the microtubule. In practise and also in our
simulations, this quantity varies in time.  In \sec{sec:poisson} we
Poisson-averaged the theoretical predictions thus rendering it
possible to compare theory and simulation results. Using the area
fraction on the cargo from which the microtubule is in binding range
for the motors as a fit parameter, we found good agreement between
theory and simulations for both the mean run length and the mean
number of bound motors. Note that the latter one cannot be measured in
typical bead assay experiments.

We also determined the mean separation height between cargo and microtubule and found
$\mean{h} = 4-14$~nm. Modeling the motor stalk as a cable resulted in
smaller distances than using a full spring model for the motor stalk. 
A recent experimental study using fluorescence
interference contrast microscopy found that kinesin holds its cargo
about 17~nm away from the MT \cite{diez:06}. Our smaller distance probably
results from neglecting any kind of volume extension (except binding site
occupation) of the motor protein, the simplified force extension relation
applied to model the stalk behavior, and neglecting electrostatic
repulsions. These effects, however, could easily be included into our algorithm,
e.\,g., by using hard core interactions that account for the finite volume of
the motor protein segments.

In \sec{sec:inrange} we explicitely demonstrated that the theoretical assumption of
having a fixed number of motors in binding range during one walk is not
justified (cf. \fig{fig:singleinrange}). Nevertheless the theoretical results
agree well with the simulations. This might be explainable by the
observation that averaged over many runs the distribution of motors in binding
range appears to be Poissonian (cf. \fig{fig:inrange}). Thus, on the one hand
fast fluctuations in the number of motors in binding range around some mean
value are not visible. On the other hand periods in which this number
fluctuates around the same mean value can be treated as a complete run. Thus,
averaging over these smaller runs (i.\,e., which end after the sphere was
e.\,g.\ rotated visibly and not after the last motor unbinds) has the same
effect as averaging over complete runs (i.\,e., which end after the last motor
unbinds).

An interesting question is to what extend several motors can cooperate
by sharing load. We have addressed this question for the case of
several motors pulling a cargo particle against high viscous friction.
One of the advantages of our algorithm is that we can measure the
velocity of the bead and at the same time also the number of
simultaneously pulling motors. Thus, in \sec{sec:velocity} we tried to
check whether the explanation of Ref.~\cite{hill:04} is correct also
under the assumptions of our model, especially for the parameter
values given in \tab{tab:parameter}.  Our simulations show that the
speed of the cargo increases with the number of pulling motors for
high viscous friction, in agreement with experimental results
\cite{Hunt__Howard1994}. Our simulations however show pronounced
deviations from the quantitative predictions based on the assumption
that the load is shared equally among the bound motors. Furthermore,
as the average life time of a state with a certain number of pulling
motors is rather short the different velocities expected for different
numbers of instantaneously pulling motors were smeared out by
diffusion.  Similarly when directly measuring how the total load force
is distributed to the different motors pulling, no equal load sharing
could be observed for the escape rate of about 1~s$^{-1}$. We observed
equal load sharing only when we used a ten-fold smaller escape rate,
in order to increase the life time of the motors in the bound state.

Another interesting question in this context is whether the velocity
distribution exhibits several maxima if the cargo is pulled against a
viscous load, as observed in several experiments {\it in vivo}
\cite{hill:04,Levi__Gelfand2006}. For example, Hill et al.
\cite{hill:04} found that vesicle in neurites move with constant
velocity for some period of time and then switch to another constant
velocity in a step-like fashion. The distribution of velocities
(measured over time intervals of the order of 1~s) was found to have
peaked at integer multiples of the minimal observed velocity. These
peaks were interpreted as corresponding to different numbers of
simultaneously pulling motors, which equally share the visoelastic
load excerted by the cytoplasm \cite{hill:04} (cf. also
\eq{eq:velratio}). Indeed, both an earlier model for motor cooperation
\cite{klumpp:05} and our present description predict that equal
sharing of a large viscous load leads to such a velocity distribution.
In our simulations, we could however not resolve multiple peaks,
presumably because the peaks are too broad to be resolved. The latter
results from a combination of the way how we measure the velocity and
from the fast dynamics of motor unbinding as discussed in
\sec{sec:velocity}.

As already mentioned above, the framework of our method is rather
general. Therefore various model variations can be easily implemented
and probed in simulations. Here, for example we modeled the motor
stalk by two versions of a simple harmonic spring: the cable model,
which represents a molecule with an intrinsic hinge, and the spring
model. More advanced force--extension relations could easily be
incorporate in \eq{eq:singlemotorforce} in order to probe the
influence of more realistic polymer models on the transport process.
Similarly, the force dependence in unbinding from the microtubule and
stepping can be altered to explore the impact onto macroscopic
observables like the mean run length or the speed of the cargo.
Furthermore, the algorithm could easily be adapted to study
beads to which clusters of motors or defined motor complexes (such as those in ref. \cite{rogers09})
are attached. Thus, the algorithm described in this paper can be regarded as a link between
purely theoretical models and data from \emph{in vitro} experiments.

Another interesting question for future applications of our method is
how cargo transport works against some external shear flow. Since our
model is based on a hydrodynamic description, flow can easily be
included in the dynamics of our model. For these studies the
Langevin-equation, \eq{langevin-ito}, has to be extended by additional
terms accounting for the effect of an incident shear flow
\cite{brady:89,korn:08a}.  Then, two opposing effects exist
characterized by the step rate and the strength of the shear flow,
respectively. Their interplay together with the rates for binding and
unbinding $\pi_{ad}$ and $\epsilon$, respectively, determine whether
the cargo moves in walking direction or in flow direction.
Experimentally, such a setup might provide interesting perspectives
for biomimetic transport in microfluidic devices.

\begin{acknowledgments}
  This work was supported by the Center for Modelling and Simulation
  in the Biosciences (BIOMS) and the Cluster of Excellence Cellular
  Networks at Heidelberg.  S.K. was supported by a fellowship from
  Deutsche Forschungsgemeinschaft (Grant No. KL818/1-1 and 1-2) and by
  the NSF through the Center for Theoretical Biological Physics (Grant
  No.  PHY-0822283).
\end{acknowledgments}

\appendix

\section{Adhesive motor dynamics}
\label{appendix:mdynamics}

The Langevin equation, \eq{langevin-euler}, and the motor dynamics rules
explained in \sec{sec:mdynamics} are connected by the following algorithmic
rules that apply in each time step $\Delta t$:
\begin{itemize}
  \item[(i)] The sphere's position and orientation is updated according to
    \eq{langevin-euler} (for an explicit description see
    Ref.~\cite{korn:07a}).
  \item[(ii)] The positions where the motors are attached to the sphere in the
  flow chamber coordinate system are calculated.
  \item[(iii)] For each motor that is bound to the MT its load force is
  calculated. Then stepping is checked according to the stepping rate derived
  from \eq{force-velocity}. If the motor steps forward the load force is again
  calculated as motor length/direction has changed.
  \item[(iv)] For each motor that is not bound to the MT binding is checked
  according to the procedure explained in the main text (\sec{sec:mdynamics}). 
  \item[(v)] For each active motor (i.\,e., bound to the MT), the contribution
  to $\boldv{F}^D$ is calculated.
  \item[(vi)] Each motor that is bound to the MT can unbind with escape rate
 $\epsilon$ given by the Bell equation, \eq{eq:bell}.
\end{itemize}
A motor that escaped from the MT in one time step can rebind to the MT in the next time step
according to rule (iv). The same Monte-Carlo technique that is explained in
the main text (\sec{sec:mdynamics}) to decide whether binding occurs or not is
also used for the decission on stepping and unbinding. For the pseudo random
number generator we used an implementation of the Mersenne Twister \cite{mersenne}.

\section{Motor distribution algorithm}
\label{appendix:distribution}

Initially the center of the sphere is located at position $(0,0,R+l_0+h_{MT})$
directly above a microtubule binding site (cf. \fig{fig:setup}). The first
motor that is distributed is initially fixed at position $(0,0,l_0+h_{MT})$
(relative to the chamber coordinate system) with its tail. The head is bound
to the microtubule at $(0,0,h_{MT})$. Thus the initial distance between the
motor and the microtubule is given by the motor resting length $l_0$.  For the
distribution of the other $N_{tot}-1$ motors on the sphere we use an hard disk
overlap algorithm similarly to the one that was described in
Ref.~\cite{hammer:92}. For each of these motors two random variables are
chosen $r_1$ from the uniform distribution on $(0, 2\pi)$ and $r_2$ from the
uniform distribution on $(0,1)$, respectively. Then, the motor is located on the sphere's surface
at the spherical coordinates $(r_1, \arccos(1-2r_2))$ and possible overlap to
already distributed motors is checked. If no other motor is located within a
ball of radius $0.1l_0$ around the just distributed motor, then its position
is kept, otherwise a new pair of random coordinates are drawn until no overlap
with other motors exists.


\end{document}